\documentclass[journal]{IEEEtran}
\usepackage[utf8]{inputenc}
\usepackage{cuted}
\usepackage[utf8]{inputenc}
\usepackage[caption=false, font=footnotesize]{subfig}
\usepackage{amsmath}
\usepackage{amsfonts}
\usepackage{booktabs}
\usepackage{colortbl}
\usepackage{multirow}
\usepackage{tikz}
\usetikzlibrary{shapes.geometric, arrows, calc}

\tikzstyle{block} = [rectangle, rounded corners, minimum width=4cm, minimum height=0.6cm,text centered, draw=black]
\tikzstyle{operation} = [circle, minimum size = 0.7cm, text centered, draw = black]
\tikzstyle{arrow} = [thick,->,>=stealth]

\title{Modulation and Signal Processing for LEO-LEO Optical Inter-satellite Links}

\author{I.~P.~Vieira~,~T.~C.~Pita~,~D.~A.~A.~Mello%
	\thanks{IPV, TCP, and DAAM are with the Department of Communications (DECOM), School of Electrical and Computer Engineering, University of Campinas (UNICAMP), Campinas, SP, 13083-852, BR (email: darli@unicamp.br).}
	\thanks{Part of this work appears in \cite{Pita2022}.}
	\thanks{This work has been supported by \textit{Idea! Electronic Systems} and by \textit{the Conselho Nacional de Desenvolvimento Cient\'{i}fico e Tecnol\'{o}gico (CNPq)}.}
}%
\date{July 2023}%

\makeatletter
\def\ps@IEEEtitlepagestyle{
  \def\@oddfoot{\mycopyrightnotice}
  \def\@evenfoot{}
}
\def\mycopyrightnotice{
  {\footnotesize
  \begin{minipage}{\textwidth}
  \centering
  Copyright~\copyright~2023 IEEE. Personal use of this material is permitted. Permission from IEEE must be obtained for all other uses, in any current or future media, including reprinting/republishing this material for advertising or promotional purposes, creating new collective works, for resale or redistribution to servers or lists, or reuse of any copyrighted component of this work in other works.
  \end{minipage}
  }
}

\begin{document}
	
	\maketitle
	
	\begin{abstract}
		We investigate key aspects of coherent optical communications on inter-satellite links (ISLs) for the next-generation ultra-dense low-Earth orbit (LEO) constellations. Initially, the suitability of QPSK, 8-QAM, and 16-QAM modulation formats with different symbol rates (28 GBaud, 60 GBaud, and 120 GBaud) and channel coding schemes (oFEC and staircase codes) for intra- and interorbital connections is evaluated. We provide SNR margins for all investigated sets and determine unfeasible operating points. We show that sets with higher-order modulation formats combined with high symbol rates can prove unfeasible, even for first-neighbor connections. Furthermore, the presence or absence of optical pre-amplification as well as the choice for a more robust channel coding technique, such as the oFEC, can be decisive in certain LEO-LEO links. Next, we characterize the Doppler shift (DS) and its time derivative for first-neighbor interorbital connections in two different topologies and for general connections established between any pairs of satellites. Our results reveal that while the maximum Doppler-generated frequency shift amplitude can be considerably higher than those typically found in fiber-optic communications, the time derivative values are significantly lower. Finally, we address all-digital DS compensation in extreme cases of frequency offset amplitude and derivative where the typical Mth-power algorithm is not sufficient. To this end, we propose a filtered version of an existing two-stage method combining spectral shifts with the Mth-power method. The simulation results indicate that this approach provides an appropriate solution for all examined cases. 
	\end{abstract}
	
	
	\section{Introduction}
	\IEEEPARstart{O}{ptical Inter-Satellite Links} (OISLs) play a critical role in helping next-generation low-Earth orbit (LEO) constellations achieve two of their chief aims: (1) global coverage and (2) low latency \cite{liu2021leo,al2022next}. Independence from immediately reachable ground gateway stations -- i.e. those belonging to the satellites' field of view (FoV) -- is guaranteed through the formation of multi-hop relaying satellite networks, providing a cost-efficient integration of polar and oceanic regions \cite{reid2018broadband,leyva2020leo}. Low latency relies on the fact that the beam propagation speed is about 50\% higher in vacuum than in the optical fiber, revealing, therefore, an advantageous alternative for path routing in ultra-long distances between source and destination \cite{handley2018delay,chaudhry2021optical,li2022delay}.
	
	Recently, private companies like Telesat and SpaceX, which own some of the largest incoming LEO constellations, have stated in their petitions to the Federal Communications Commission (FCC) that they plan to use inter-satellite links (ISLs) in their systems \cite{del2019technical,pachlerupdated}.
	Deciding for optical carriers rather than microwave or the conventional radio-frequency (RF) ones when establishing ISLs is justified by a coalition of factors suited to the new communication technologies, including high bandwidth, high beam selectivity, reduced power consumption, smaller receiver's telescopes, and improved security \cite{kaushal2016optical,leyva2021inter,yue2022security,toyoshima2022applicability}. Furthermore, modern techniques in digital coherent optics allow operation at high data-rate regimes over long distances owing to high receiver sensitivity, high-order modulation schemes, and anti-interference ability \cite{li2022survey,yu2020digital}.
	
	FSO and fiber-optic communication (FOC) systems are both laser-based technologies that differ essentially by the beam’s propagation medium and its corresponding impairments, retaining most of the transceiver components. Modern hollow-core fibers \cite{klaus2022hollow,richardson2022hollow} may narrow the gap even more. This closeness is very opportune as it enables the reuse of techniques in digital signal processing (DSP) as well as the device manufacturing expertise from wired terrestrial network infrastructure. While LEO-LEO OISLs are subject to neither non-linear effects -- which would occur in the presence of the glass waveguide \cite{singh2007nonlinear} -- nor degradations resulting from atmospheric phenomena \cite{almonacil2021experimental,allain2021enhanced,walsh2022demonstration}, such links pose very specific challenges to their implementation. 
	
	In this work, we cover two of the major channel impairments for LEO-LEO OISLs in the opto-electronic domain. Initially, we examine the high free-space path loss (FSPL) resulting from the geometric scattering of the beam in the vacuum. Here, the main interest is to measure the performance of different modulation formats and data rates for these links, assuming the use of typical forward error correction code (FEC) classes -- namely, staircase codes, as per the standard ITU-T G.709.2/Y.1331.2 \cite{ITU-T}, and open FEC (oFEC), presented in the OpenZR+ MSA technical specification \cite{OpenZR}. Next, we focus on the Doppler-generated frequency offset induced by the relative motion between the transmitter (Tx) and receiver (Rx) terminals. This frequency offset is especially important when addressing satellites at low altitudes since the relative velocities involved are far above those observed for satellites in medium and geosynchronous orbits. Beyond the scope of this work, there are a number of relevant issues for LEO-LEO interconnection that have been extensively investigated in the literature, including point-ahead-angle estimation, acquisition and tracking accuracy, and satellite vibration \cite{kaushal2017free}.
	
	The remainder of the paper is organized as follows. First, in Section II, we provide a brief overview of related works. In Section III, we introduce the Walker constellation model (WCM) and constellations architectures. In Section IV, we study the suitability of different modulation formats and data rates for first-neighbor intra- and interorbital coherent communication. The subsequent sections are focused on Doppler shift: its characterization in the context of LEO-LEO OISLs is presented throughout Section V, whereas an all-digital methology for its compensation is described in Section VI. In Section VII, we show simulation results assessing bandwidth requirements and compensation capabilities for different modulation formats. Finally, Section VIII concludes the paper.
	
	\section{Related Works}

    Over the past two decades, a number of successful in-orbit demonstrations have gradually proven the potential of FSO technology to spearhead the next generation of space telecommunication systems. Institutions like the European Space Agency (ESA), the \textit{Deutsches Zentrum für Luft- und Raumfahrt} (DLR), the Japan Aerospace Exploration Agency (JAXA), and Tesat-Spacecom (TESAT) have provided important contributions to the experimental domain of LEO-LEO OISLs \cite{sodnik2010optical,hemmati2020near}. Meanwhile, other players like the National Aeronautics and Space Administration (NASA, USA) and the National Institute of Information and Communications Technology (NICT, Japan) have concentrated their efforts predominantly on the development of the satellite-to-ground optical links \cite{li2022survey,hemmati2020near}. Some of the most remarkable achievements from the first era of the space-to-space applications, which takes place from the 90s to the mid-aughts, were the Communications Research Laboratory (CLR)'s ETS-VI/LCE \cite{arimoto1995preliminary}, ESA's SILEX \cite{tolker2002orbit}, and JAXA's LUCE \cite{nakagawa1995preliminary}, which were able to guarantee rates of tens or hundreds of Mbps in LEO-to-geostationary (LEO-GEO) connections \cite{zhang2023laser}. The choice for on-off keying (OOK) and pulse-position modulation (PPM), with intensity modulation (IM), in these early stages, can be attributed to its low-complexity receiver mechanisms and the maturity of this technology at the time \cite{li2022survey,ke2023coherent}. On the other hand, great attention has been devoted to the use of coherent optics in ISLs from the success of TESAT/DLR's NFIRE-to-TerraSAR-X \cite{fields2009nfire} (LEO-LEO at 5.6 Gbps over link distance of up to 5000 km) and ESA's EDRS-A \cite{poncet2017hosting} (LEO-GEO at 1.8 Gbps over link distance of up to 45,000 km), delivered in 2008 and 2016, respectively. In these missions, it was possible to achieve rates in the range of Gbps by way of  BPSK modulation \cite{ke2023coherent}. Despite their inherent complexity, coherent schemes are an attractive pick to ensure adequate background noise rejection and enhanced spectral efficiency when compared to IM/DD ones \cite{khalighi2014survey}. In a recent work, Guiomar et al. \cite{guiomar2021400g+} demonstrate the reliability of outdoor FSO transmission supporting more than 400 Gbps per channel in the presence of atmospheric turbulence and weather conditions for a short-range link. A NICT team is currently developing miniaturized space laser-communication terminals for general purpose that allow LEO-LEO full-duplex communication up to 100 Gbps \cite{carrasco2022free}. We refer the interested reader to \cite[Chapter 2]{zhang2023laser} for a very detailed timeline of the OISLs evolution.
	
	The feasibility of a certain digital modulation format in a communication link is closely related to the system's bit error rate (BER) performance. In general, there is a trade-off between the increase in spectral efficiency, promoted by higher-order modulation formats, and the transmission reach. In \cite{vieira2022link} we quantify typical inter-satellite distances and their corresponding FSPLs for first-neighbor type connections for four of the largest emerging LEO constellations, highlighting the importance of the phase factor for connections between satellites belonging to adjacent orbital planes. Li et al. \cite{li2022survey} provide some potential applications for different modulation formats, considering both coherent and non-coherent detection schemes, for various space-based communication networks, including LEO-LEO connections. Liang et al. \cite{liang2022link} investigate the link margin on LEO OISLs assuming on-off keying (OOK) as modulation scheme. The relationship between BER and received power for OISLs using intensity modulation and direct detection (IM/DD) is also studied by Carrizo et al. \cite{carrizo2020optical} on the LEO architectures described in \cite{del2019technical}. Maho et al. \cite{maho2019assessment} perform a comparative study of the performance for systems based on differential phase-shift keying (DPSK) and OOK schemes that include LEO downlinks and LEO-LEO OISLs up to 10 Gbps. Few studies, however, are dedicated to the use of amplitude and phase modulated/coherent detection schemes in OISLs, the majority being focused on ground-to-satellite connections \cite{hemmati2021advances}. This work aims to fill this gap through a detailed study of the suitability of different modulation formats in OISLs for first-neighbor connections (FNCs) in some of the largest next-generation LEO constellations, taking into account different symbol rates and the use of channel coding techniques.
	
	The second part of the paper is dedicated to the characterization and compensation of DS in ultra-dense LEO constellations. The presence of DS in FSO communications is a widely documented phenomenon \cite{majumdar2022laser,tan2022effect}. However, to the best of our knowledge, there is no comprehensive database available in the literature on the typical levels of DS for OISLs in the LEO constellations context. So, as a lead-in task, we perform the characterization of DS frequency offset and its time derivative, taking into account four benchmarking constellation layouts -- the same ones mentioned in the previous paragraph. Next, we move on to the compensation strategy. Few proposals in the satellite communication literature have addressed optical Doppler shift compensation (DSC). Most previous works propose to carry out optical domain frequency-shift compensation using optical phase-locked loops (OPLLs). Ando et al. employ a Costas loop for DSC \cite{Ando2011}. Yue et al. implement a decision-driven scheme with digital filters \cite{Yue2018}. Liu et al. propose a multistage back loop for a wider compensation range \cite{Liu2018}. Schaefer et al. suggest a loop filter with adjustable gain to mitigate receiver power variations \cite{Schaefer2015}. Leveraging the maturity of digital coherent optical systems, a natural step is to perform DSC in the digital domain with DSP-based frequency offset algorithms originally designed for fiber-optic systems. From this perspective, Almonacil et al. propose in \cite{Sylvain2020} a DSP-based DS transmitter pre-compensation scheme, avoiding the need for excess receiver bandwidths. In \cite{Pita2022} we evaluate the DSC performance achieved by an existing frequency offset post-compensation technique due to Diniz et al. \cite{Diniz2011} based on power spectrum imbalances. However, such analysis does not apply to the current scenario under investigation as it focuses on frequency offsets with amplitudes lower than 5 GHz (typically, the highest value used in FOCs), non-return-to-zero (NRZ) and return-to-zero (RZ) pulse shapes, and narrow bandwidths. Conversely, we are interested in frequency offsets of up to 10 GHz in systems with Nyquist pulse shaping and with extended excess bandwidth, bringing out the need for additional filtering stages. 
	
	
	\section{Orbit Architecture}
	\subsection{System Model}
\begin{figure}[t]
		\centering
		\includegraphics[width=\columnwidth]{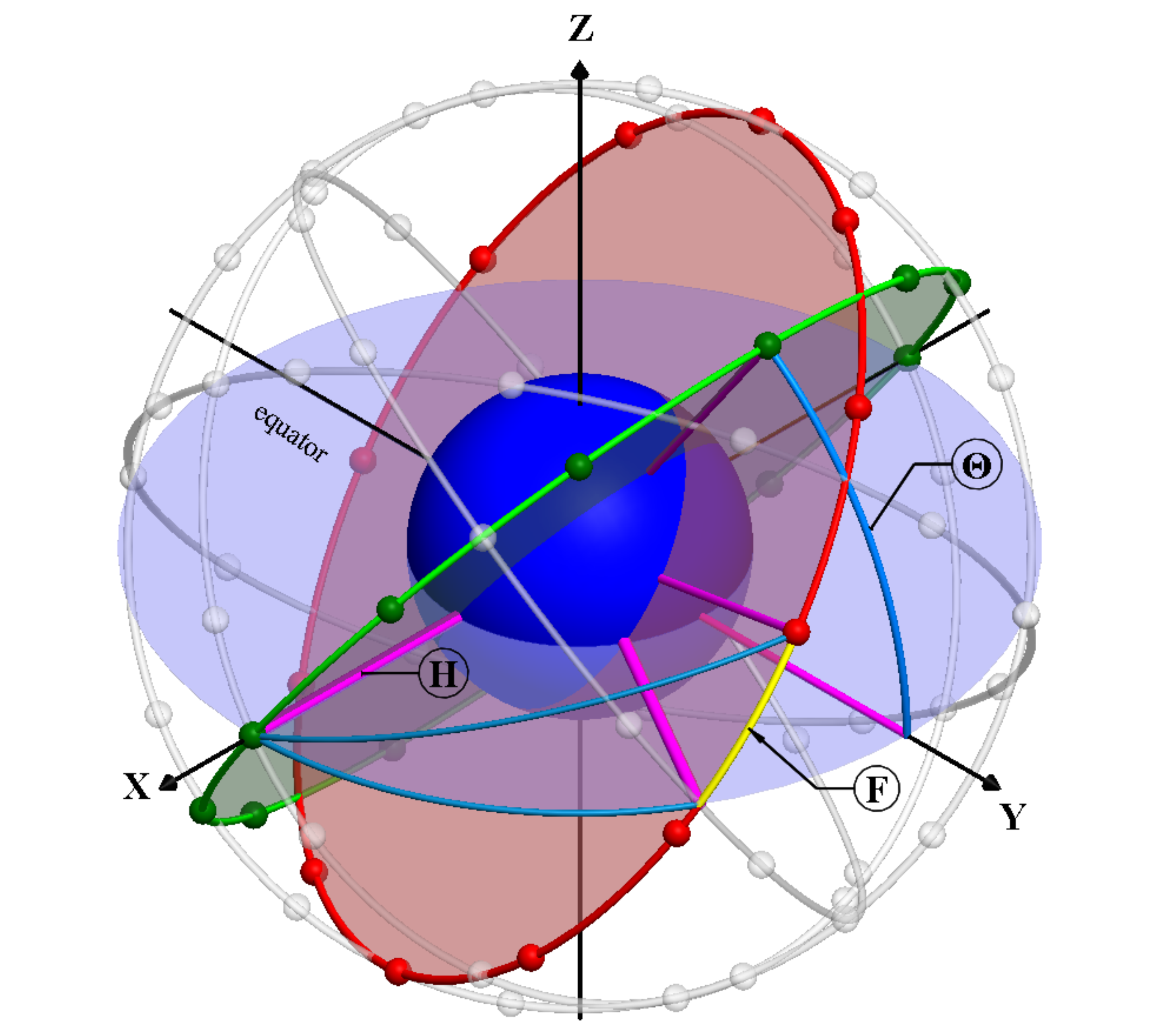}
		\caption{Walker constellation parameters. Two adjacent planes highlighted (in red and green colors). The small spheres evenly distributed along these planes represent the satellites. $H$ denotes the constellation's altitude with respect to the Earth's surface. $\theta$ is the orbital plane's inclination angle, common to all planes. Finally, $F$ is the phase factor, which corresponds to the plane-to-plane relative angular offset suffered by the k-th satellites with reference to the equatorial/ecliptical plane.}
		\label{fig:wcm}
	\end{figure}

	Let $r_{ik}^{\xi}(t)=x_{ik}\hat{x}+y_{ik}\hat{y}+z_{ik}\hat{z}$ be the position vector for the $k$-th satellite in the $i$-th circular orbital plane, with $k\in\left[\right.0,S\left.\right)$ and $i\in\left[\right.0,P\left.\right)$, where $S$ is the number of satellites per plane and $P$ denotes the total of planes. The superscript $\xi\in\left\{s,d\right\}$ is used for distinguishing between source ($s$) and destination ($d$) satellites. According to the WCM \cite{walker1984satellite},
	\begin{equation}
		\left[\begin{array}{c}
			x_{ik}\\
			y_{ik}\\
			z_{ik}
		\end{array}\right]=R\left[\begin{array}{c}
			\cos\theta\sin\rho\sin\Omega(t)+\cos\rho\cos\Omega(t)\\
			\cos\theta\cos\rho\sin\Omega(t)+\sin\rho\cos\Omega(t)\\
			\sin\theta\sin\Omega(t)
		\end{array}
		\right],
		\label{eq:Model}
	\end{equation}
	for
	\begin{align*}
		\rho = \left(\frac{2\pi i}{P}\right) &&
		\textrm{and} &&
		\Omega(t)=\omega t+2\pi\left(\frac{k}{S}+\frac{iF}{PS}\right),
	\end{align*}
	where $R$ is the constellation's altitude, $H$, plus the Earth's radius ($R_\oplus\approx 6.371\textrm{ km}$), $\theta$ is the (common) inclination of the orbital planes, $\omega $ is the angular velocity of the satellites, $F$ is the phase factor between satellites belonging to adjacent orbital planes -- as schematized in Fig. \ref{fig:wcm} --, and $t$ is the time. This model is frequently used in many constellation design projects due to its regularity: both the satellites inside the planes as well as the planes around the globe are uniformly distributed -- the former being $2\pi/S$ rad apart, while the latter, $2\pi/P$ rad --, resulting in a highly symmetrical constellation. Once the altitude and inclination are established, a Walker constellation can be uniquely identified by the triplet ``$\theta$: N/P/F'', called Walker notation, where $N=P\times S$ refers to the total number of satellites.
	
\begin{figure*}[ht!]
		\centering
		\subfloat[$k$-to-$k$.\label{k2k}\label{fig:connection-a}]{%
			\includegraphics[width=.45\textwidth]{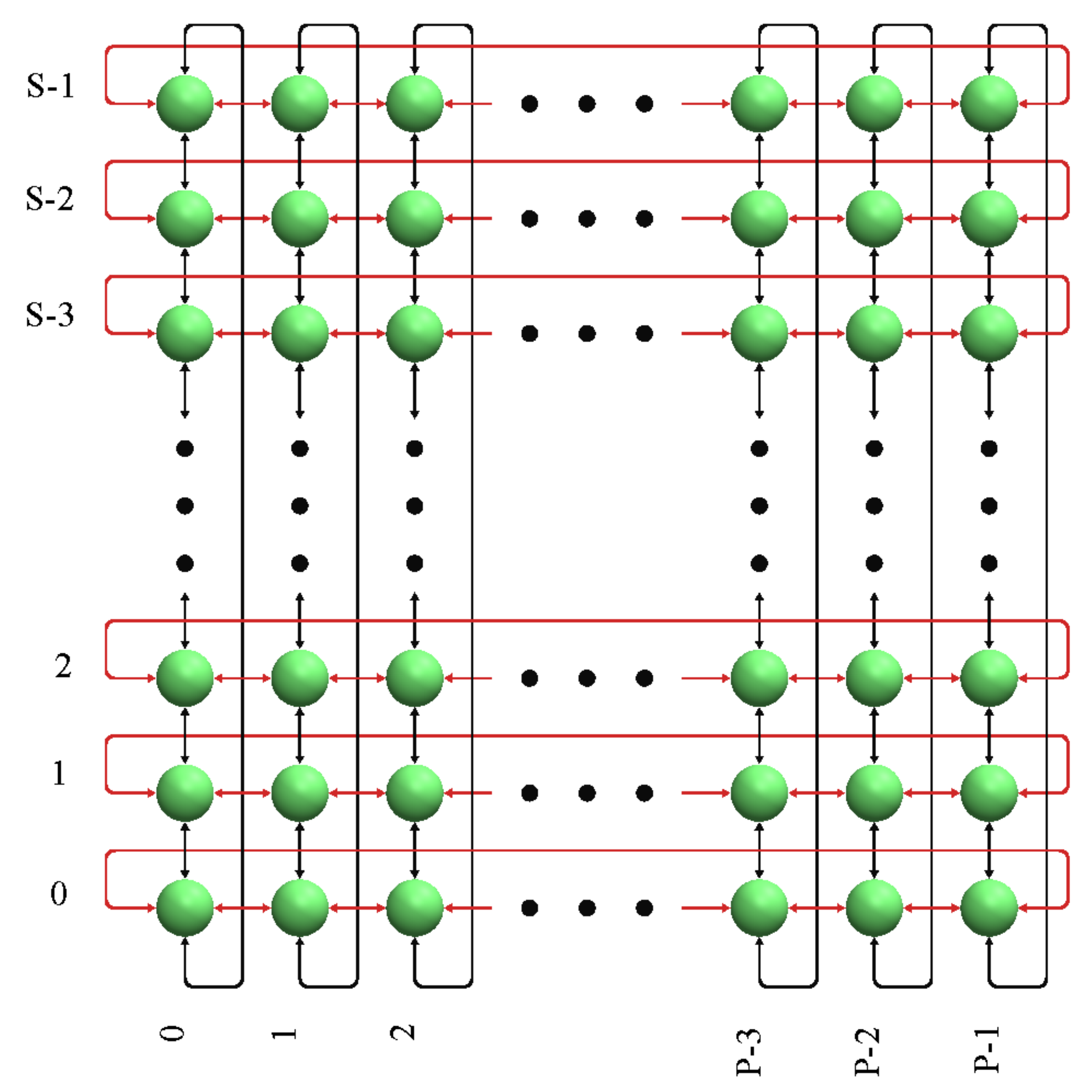}}
		\qquad
		\subfloat[$k$-to-$k-1$.\label{k2k-1}\label{fig:connection-b}]{%
			\includegraphics[width=.45\textwidth]{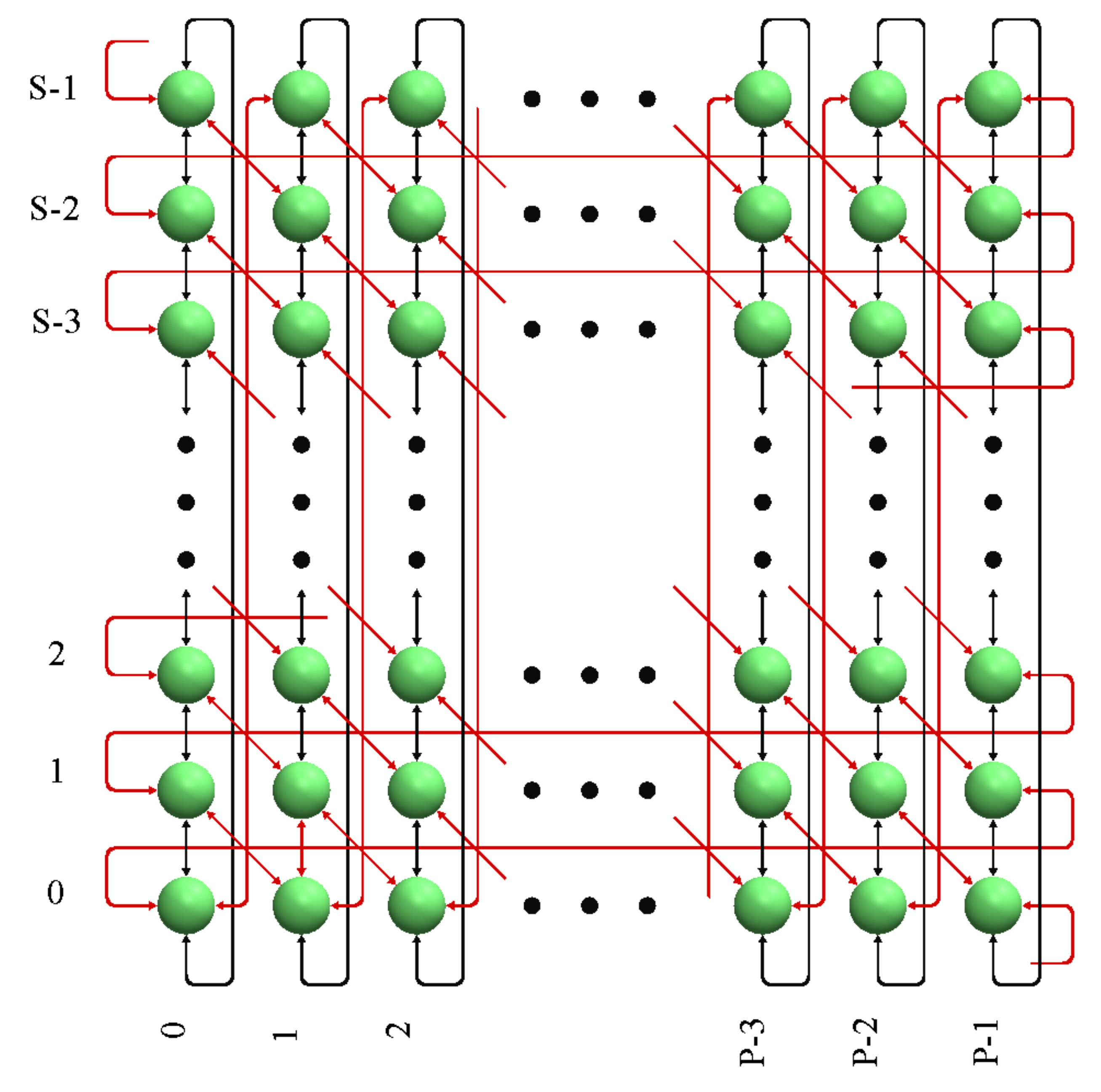}}
		\caption{First-neighbor connection patterns (illustrated for $F=0$). Current networking architectures for large LEO constellations consider the presence of four satellite first-neighbor connections, among which two are established with satellites in the same orbital plane (drawn in black) and the other two with satellites in adjacent planes (drawn in red). The Doppler shift in intraorbital links is null given the absence of eccentricity in the satellites' trajectory in the Walker constellation model's circular orbit approximation. Interorbital connections, in turn, have non-negligible Doppler shift values, and are assessed here in two topologies: (a) $k$-to-$k$ and (b) $k$-to-$k-1$.}
	\end{figure*}

	OISLs can be classified into two categories broken down by the nature of the connection, called first-neighbor and all-to-all connections. In the context of FNCs, four possible links are evaluated \cite{yang2009doppler}. Two of them have intraorbital nature, connecting $k$-to-$k+1$ and $k$-to-$k-1$ satellites in the same orbital plane (i.e., keeping the $i$ parameter fixed). The other two have interorbital nature, connecting satellites $k$-to-$k$ or $k$-to-$k-1$ in adjacent planes. Interorbital OISLs appear in red in Fig. \ref{fig:connection-a} ($k$-to-$k$) and Fig. \ref{fig:connection-b} ($k$-to-$k-1$). Unlike FNCs, the all-to-all connections (AACs) are those that are established between any pair of satellites along the entire shell, with FNCs as a particular case. In this type of connection, continued communication between the source and destination satellites is commonly precluded, either by Earth occlusion or by a physical limitation regarding the link's reach \cite{chaudhry2022temporary}.
	
	\subsection{Commercial constellations}
	We evaluate typical parameters currently practiced by four of the largest companies in the LEO-based broadband sector \cite{del2019technical,pachlerupdated}, labeled here from \textbf{A} to \textbf{D}. Such constellations are at different stages of maturity. In order to ensure a common basis for comparison, we take them in their final versions (full deployment), as reported in FCC filings as of January 2021 \cite{tsfcc1,tsfcc2,owfcc1,owfcc2,owfcc3,sxfcc1,sxfcc2,sxfcc3,sxfcc4,azfcc1}. Table \ref{table:pachler} shows a summary of the orbital characteristics of each of them. 
	
	\begin{table}[ht!]
		\centering
		\caption{Orbit characteristics of fully deployed LEO constellations. Q$^{(m)}$ stands for the $m$-th shell of the system $\mathbf{Q}\in\left\{\mathbf{A},\mathbf{B},\mathbf{C},\mathbf{D}\right\}$. \cite{pachlerupdated}}
		\label{table:pachler}
		\begin{tabular}{>{\centering}m{0.05\textwidth}>{\centering}m{0.05\textwidth}>{\centering}m{0.05\textwidth}>{\centering}m{0.04\textwidth}>{\centering}m{0.04\textwidth}>{\centering}m{0.04\textwidth}>{\centering\arraybackslash}m{0.04\textwidth}}
			System & Shell & $H$ [km]
			& $\theta$ [\textdegree] & $P$ & $S$ & $N$ \\
			\midrule
			\multirow{2}{*}{\textbf{A}} & \cellcolor{blue!10}A$^\text{(1)}$ &\cellcolor{blue!10}1,015 &\cellcolor{blue!10}98.98 &\cellcolor{blue!10}27 &\cellcolor{blue!10}13 & \multirow{2}{*}{1,671} \\
			& A$^\text{(2)}$ & 1,325 & 50.88 & 40 & 33 &\\
			\midrule
			\multirow{3}{*}{\textbf{B}} & \cellcolor{blue!10}B$^\text{(1)}$ & \cellcolor{blue!10}1,200 &\cellcolor{blue!10}87.9 &\cellcolor{blue!10}36 &\cellcolor{blue!10}49 & \multirow{3}{*}{6,372}\\
			& B$^\text{(2)}$ & 1,200 & 55 & 32 & 72 & \\
			& \cellcolor{blue!10}B$^\text{(3)}$ & \cellcolor{blue!10}1,200 &\cellcolor{blue!10}40 &\cellcolor{blue!10}32 &\cellcolor{blue!10}72 &\\
			\midrule
			\multirow{5}{*}{\textbf{C}} & C$^\text{(1)}$ & 540 & 53.2 & 72 & 22 & \multirow{5}{*}{4,408}\\
			& \cellcolor{blue!10}C$^\text{(2)}$ & \cellcolor{blue!10}550 & \cellcolor{blue!10}53 &\cellcolor{blue!10}72 & \cellcolor{blue!10}22 &\\
			& C$^\text{(3)}$ & 560 & 97.6 & 6 & 58 &\\
			& \cellcolor{blue!10}C$^\text{(4)}$ & \cellcolor{blue!10}560 & \cellcolor{blue!10}97.6 & \cellcolor{blue!10}4 & \cellcolor{blue!10}43 &\\
			& C$^\text{(5)}$ & 570 & 70 & 36 & 20 &\\
			\midrule
			\multirow{3}{*}{\textbf{D}} & \cellcolor{blue!10}D$^\text{(1)}$ & \cellcolor{blue!10}590 & \cellcolor{blue!10}33 & \cellcolor{blue!10}28 & \cellcolor{blue!10}28 & \multirow{3}{*}{3,236}\\
			& D$^\text{(2)}$ & 610 & 42 & 36 & 36 &\\
			& \cellcolor{blue!10}D$^\text{(3)}$ & \cellcolor{blue!10}630 & \cellcolor{blue!10}51.9 & \cellcolor{blue!10}34 & \cellcolor{blue!10}34 &\\
			\bottomrule
		\end{tabular}
	\end{table}
	
	All these constellations have in common the use of a huge number of satellites with small orbital periods, ranging from 95 min (for C$^{(1)}$, at $540$ km) to 112 min (for A$^{( 2)}$, at $1,325$ km). There is, however, no clear pattern in the choice of orbital characteristics. System \textbf{B}, which has the largest number of satellites among all, is the only one to keep all shells at the same altitude ($1,200$ km). Some systems, like \textbf{A} and \textbf{C}, choose to make the number of orbital planes greater than the number of satellites per plane, unlike \textbf{B}, which invests in a higher orbital density and fewer planes. System \textbf{D}, in turn, adopts square shells, where the number of orbital planes is always equal to the number of satellites per plane. Furthermore, with the exception of \textbf{D}, which has satellites only in medium-inclination orbits (coincident with the most densely populated areas on Earth \cite{pachlerupdated}), all the others systems allocate a small portion of their total satellite capacity, about 10\% to 30\%, in polar orbits.
	
	\section{Modulation formats for commercial LEO ISLs}
	
	In this section, we evaluate the suitability of different modulation formats/data rates for the previous constellation architectures. For this purpose, the shot-noise- and ASE-limited SNR values are determined for intra- and interorbital FNCs, assuming $k$-to-$k$ and $k$-to-$k-1$ topologies.  The analysis is based on the pre-FEC BER of two typical coding schemes. The staircase FEC code defined in ITU-T G.709.2/Y.1331.2 requires approximately a $4.5\cdot 10^{-3}$ pre-FEC BER \cite{ITU-T}, whereas the oFEC defined in the OpenZR+ MSA specification requires $2\cdot 10^{-2}$ \cite{OpenZR}.
	
	For square QAM constellations, the BER distributions as a function of the per-polarization SNR value, assuming an additive white Gaussian noise (AWGN) channel, can be obtained through the approximation \cite{mello2021}
	\begin{equation}
		\text{BER}_\textrm{sq}(\textrm{SNR}) \approx \frac{1}{b}\left[2\left(1-\frac{1}{\sqrt{\bar{M}}}\right)\text{erfc}\left(\sqrt{\frac{3\cdot \textrm{SNR}}{2(\bar{M}-1)}}\right)\right],
		\label{ber16qam}
	\end{equation}
	where $b$ is the number of conveyed bits per symbol, $\bar{M}$ is the number of symbols of the chosen square constellation, and $\text{erfc}\left(\bullet\right)$ denotes the complementary error function \cite{temme1996special}. On the other hand the following relation can be used for the star 8-QAM constellation (via Smith approximation \cite{Smith1975}):
	\begin{equation}
		\text{BER}_\textrm{st}(\textrm{SNR}) \approx \frac{5}{4}\mathbb{Q}\left(\sqrt{  \frac{6\cdot\textrm{SNR}}{b\left(3+\sqrt{3}\right)}}\right),
		\label{ber8qam}
	\end{equation}
	where $\mathbb{Q}\left(\bullet\right)$ is the Q-function \cite{temme1996special}. 
	
\begin{figure}[t]
		\centering
		\subfloat[\label{fig:ber}]{{\includegraphics[width=\columnwidth]{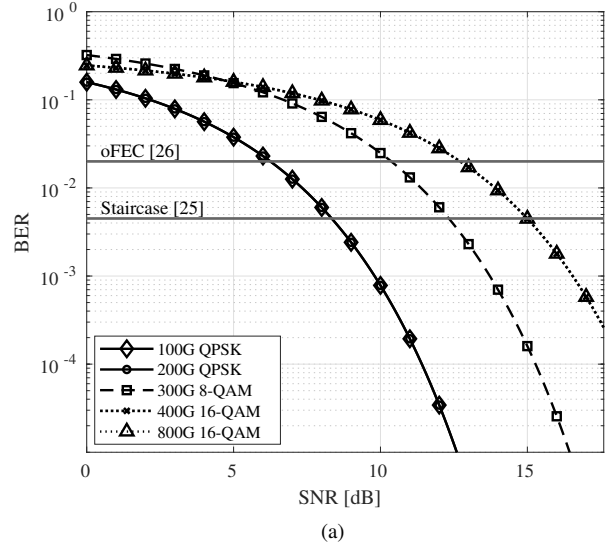}}}%
		\quad
		\subfloat[\label{fig:snr}]{{\includegraphics[width=\columnwidth]{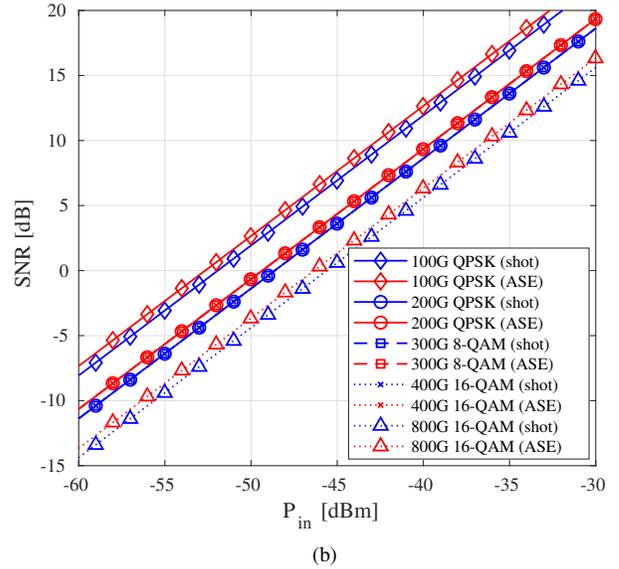}}}%
		\caption{(a) BER as a function of the SNR. (b) SNR as a function of the per-polarization incident power $P_{in}$.}%
		\label{fig:snrber}%
	\end{figure}

		The shot-noise- and ASE-limited SNR for a homodyne polarization-diversity dual quadrature coherent receiver \cite{mello2021} can be evaluated using
	\begin{equation}
		\text{SNR}_\text{shot} = \eta \langle n_s\rangle
		\label{shotdualpol}
	\end{equation}
	and
	\begin{equation}
		\text{SNR}_\text{ASE} = \frac{\langle n_s\rangle}{n_{sp}},
		\label{asedualpol}    
	\end{equation}
	where 
	 $\langle n_s\rangle \equiv P_{in}/h\nu R_s$, is the average number of photons received per symbol, $P_{in}$ is the per-polarization incident power (corresponding to half the total average power on the receiver antenna), $R_s$ is the symbol rate, $h \approx 6.62607004\cdot 10^{-34} \text{ m}^2 \text{ kg} / \text{s}$ is the Planck's constant, $\nu\equiv c/\lambda_s$ is the carrier frequency, $\lambda_s$ is the carrier wavelength, and $c \approx 3\cdot 10^8$ m/s is the speed of light in vacuum \cite{ip2008coherent}. Table \ref{tab:symbol} shows the different values assumed by $R_s$ depending on the chosen setup. In Eq. \eqref{shotdualpol}, $\eta$ is the quantum efficiency while in Eq. \eqref{asedualpol}, $n_{sp}$ denotes the spontaneous emission noise factor of the amplifier (approximated as half the amplifier noise figure, i.e. $n_{sp}\approx F_n/2$). In our simulations, $\eta = 0.7$ A/W and $F_n = 4.8$ dB. Fig. \ref{fig:ber} shows the BER as a function of the SNR for different modulation schemes/data rates. The pre-FEC BER threshold values assumed for the staircase and the oFEC are highlighted. Fig. \ref{fig:snr} shows the SNR vs. $P_{in}$ curves for shot-noise- and ASE-limited regimes at different data rates/modulation formats.

    \begin{table}[t]
		\centering
		\caption{Symbol rate values assigned to the different setups.}
		\label{tab:symbol}
		\begin{tabular}{|>{\centering}m{0.55\columnwidth}|>{\centering\arraybackslash}m{0.35\columnwidth}|}
			\hline
			100G QPSK & $28$ GBaud \\
			\hline
			200G QPSK & \multirow{3}{*}{$60$ GBaud}\\\cline{1-1}
			300G 8-QAM & \\\cline{1-1}
			400G 16-QAM & \\
            \hline
            800G 16-QAM & $120$ GBaud \\
			\hline
		\end{tabular}
	\end{table}
 
	Following \cite{hemmati2020near}, the link budget expression for the optical communication channel is given by
	\begin{equation}
		2P_{in} = P_t\tau_tG_tLG_r\tau_r\tau_jL_j,
		\label{receivedpower}
	\end{equation}
	where:
	\begin{itemize}
		\item $P_t$ is the average transmit power;
		\item $G_t = 8/w_0^2$ is the peak transmit antenna gain, which depends on the divergence angle of the Gaussian-shaped laser beam $w_0^2$;
		\item $L$ is the FSPL;
		\item $G_r = \left(\pi D_r/\lambda_c\right)^2$ is the receiving antenna gain, which depends on the Rx antenna diameter $D_r$;
		\item $\tau_{t}$ ($\tau_{r}$) is the optical loss of the transmitter (receiver);
		\item $\tau_j$ is the average pointing-loss due to random-pointing jitter;
		\item $L_j$ is the power penalty of the optical receiver caused by pointing jitter.
	\end{itemize}

		Table \ref{tab:parameters} presents typical values assumed by these parameters for OISLs and which are used in our simulations. The FSPL is a function of the link length and, therefore, of the constellation phase factor $F$. Since $F$ is not publicly available neither in FCC filings nor in the literature, we always assume the worst-case scenario, i.e., the value of $F$ that leads to the longest link length in each of the constellation architectures \cite{vieira2022link}.

	\begin{table}[t]
		\centering
		\caption{Parameter values for OISLs.}
		\label{tab:parameters}
		\begin{tabular}{|>{\centering}m{0.55\columnwidth}|>{\centering\arraybackslash}m{0.35\columnwidth}|}
			\hline
			Tx Power, $P_t$ & $1$ W \\
			\hline
			Tx optics loss, $\tau_t$ & $-2$ dB \\
			\hline
			Pointing jitter, $\sigma$ & $2.6\cdot 10^{-6}$ rad \\
			\hline
			Beam divergence half-angle, $w_0$ & $20.4\cdot 10^{-6}$ rad\\
			\hline
			Carrier wavelength, $\lambda_s$ & $1550\cdot 10^{-9}$ m \\
			\hline
			Average pointing loss, $\tau_j$ & $-0.1$ dB \\
			\hline
			Rx antenna diameter, $D_r$ & $0.1$ m \\
			\hline
			Rx optics loss, $\tau_r$ & $-2$ dB \\
			\hline
		\end{tabular}
	\end{table}
 
	\begin{figure}[t]
		\centering
		\includegraphics[width=\columnwidth]{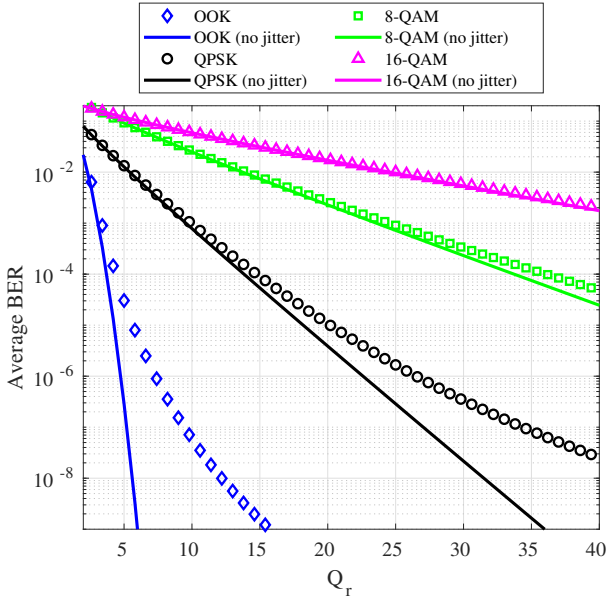}
		\caption{Average BER in presence of pointing jitter. Pointing-jitter losses become negligible when using the pre-FEC BER levels currently adopted in digital coherent optical systems, unlike the behavior observed in the OOK IM/DD scenario.}
		\label{fig:avgBER}
	\end{figure}
	
	The presence of pointing jitter tends to degrade the BER through fluctuations in the received power, thus promoting loss of system performance. In this case, the BER must be averaged with respect to the probability density function (PDF) that represents the optical intensity received by the photodetector, $p(I)$ \cite{hemmati2020near}. So,
	
	\begin{equation}
		\overline{\textrm{BER}}(Q_r) = \int_0^1 p(I)\textrm{BER}\left(IQ_r\frac{\beta+1}{\beta}\right)dI,
	\end{equation}
	with
	\begin{equation}
		p(I) = \beta I^{\beta-1},\textrm{ with }0\leq I\leq 1,
	\end{equation}
	where $Q_r$ is the required SNR parameter for the desired average BER that equals $a_\text{BER}$, $I$ is the normalized intensity, and $\beta\equiv w_0^2/(4\sigma^2)$ \cite{hemmati2020near}. This model assumes the existence of an effective control mechanism designed to track and correct constant pointing errors \cite{kiasaleh1994probability}, so that motion-induced jitter for a stable satellite system should only correspond to a small portion of the SNR budget. Its influence on system performance can be computed through the power penalty $L_j$ as
	
	\begin{equation}
		L_j =\left(\frac{Q\lvert_{\textrm{BER}(Q)=a_\textrm{BER}}}{Q\lvert_{\overline{\textrm{BER}}(Q)=a_\textrm{BER}}}\right).
	\end{equation}
	
	\noindent Fig. \ref{fig:avgBER} illustrates the dependency between the average BER, $\overline{\text{BER}}$, and the required SNR parameter. Different $w_0/\sigma$ ratios are considered, as well as different modulation schemes. The curves indicate that increasing the modulation order makes the system more robust to the presence of the pointing jitter. This behavior can be understood by observing the curves of BER as a function of the SNR, where higher-order modulation formats, such as 16-QAM, present a lower variation rate than lower-order formats, such as QPSK, especially in small SNR regions. With $w_0/\sigma = 7.89$ and $\text{BER} = 1\cdot 10^{-10}$, e.g., we have $L_j = -1.85$ dB for an OOK system with IM/DD \cite{hemmati2020near}, $L_j = -0.59$ dB for the QPSK, $L_j = -0.14$ dB for the 8-QAM, and $L_j = 0$ dB for the 16-QAM. Our feasibility analysis considers significantly higher pre-FEC BER values and only linear phase-modulated OISLs, so the $L_j$ penalty becomes negligible.

 The color scheme in Tables \ref{tab:tabshot} and \ref{tab:tabase} indicates the suitability of a given modulation scheme and the associated symbol rates for each of the shells in the shot-noise- and ASE-limited regimes, respectively: in green are those in which the use of staircase codes is enough to guarantee a post-FEC error-free transmission; in yellow, the architectures that demand the use of oFEC for such; in red, in turn, are the configurations in which the BER is greater than the oFEC threshold and therefore neither of the two coding schemes yield sufficient performance.

	\begin{table*}[ht!]
		\centering
		\caption{Suitability of different modulation formats/data rates for the various FNCs in the shot-noise-limited regime. The numbers in sequence within each cell indicate the margin, in dB, for using Staircase and oFEC as correction methods, respectively.}
		\label{tab:tabshot}
		\begin{tabular}{>{\centering}m{0.01\textwidth}>{\centering}m{0.055\textwidth}|>{\centering}m{0.14\textwidth}|>{\centering}m{0.14\textwidth}|>{\centering}m{0.145\textwidth}|>{\centering}m{0.157\textwidth}|>{\centering\arraybackslash}m{0.157\textwidth}|}
			&\multicolumn{1}{c}{}&\multicolumn{5}{c}{\textbf{Modulation Format/Data Rate}}\\
			&\multicolumn{1}{c}{}&\multicolumn{1}{c}{\textbf{100G QPSK}}
			&\multicolumn{1}{c}{\textbf{200G QPSK}}
			&\multicolumn{1}{c}{\textbf{300G 8-QAM}} & \multicolumn{1}{c}{\textbf{400G 16-QAM}} & \multicolumn{1}{c}{\textbf{800G 16-QAM}}\\
			&\multicolumn{1}{c}{}&\multicolumn{5}{c}{\cellcolor{gray!25}\textbf{INTRAORBITAL}}\\
			\cline{3-7}
			\multicolumn{1}{c}{\multirow{13}{*}{\rotatebox{90}{\textbf{ }}}}
			&\textbf{A$^{(1)}$} & \cellcolor{green!25}6.41, 8.50 & \cellcolor{green!25}3.1, 5.19 & \cellcolor{yellow!40}-0.88, 1.08 & \cellcolor{red!40}-3.55, -1.27 & \cellcolor{red!40}-6.56, -4.28\\
			\cline{3-7}
			&\textbf{A$^{(2)}$} &\cellcolor{green!25}14.07, 16.16 & \cellcolor{green!25}10.76, 12.85 & \cellcolor{green!25}6.78, 8.74 & \cellcolor{green!25}4.11, 6.39 & \cellcolor{green!25}1.10, 3.38\\
			\cline{3-7}
			&\textbf{B$^{(1)}$} &\cellcolor{green!25}17.64, 19.73 & \cellcolor{green!25}14.33, 16.42 & \cellcolor{green!25}10.35, 12.31 & \cellcolor{green!25}7.68, 9.96 & \cellcolor{green!25}4.67, 6.95\\
			\cline{3-7}
			&\textbf{B$^{(2)}$} &\cellcolor{green!25}20.98, 23.07 & \cellcolor{green!25}17.67, 19.76 & \cellcolor{green!25}13.69, 15.65 & \cellcolor{green!25}11.02, 13.30 & \cellcolor{green!25}8.01, 10.29\\
			\cline{3-7}
			&\textbf{B$^{(3)}$} &\cellcolor{green!25}20.98, 23.07 & \cellcolor{green!25}17.67, 19.76 & \cellcolor{green!25}13.69, 15.65 & \cellcolor{green!25}11.02, 13.30 & \cellcolor{green!25}8.01, 10.29\\
			\cline{3-7}
			&\textbf{C$^{(1)}$} &\cellcolor{green!25}11.50, 13.59 & \cellcolor{green!25}8.19, 10.28 & \cellcolor{green!25}4.21, 6.17 & \cellcolor{green!25}1.54, 3.82 & \cellcolor{yellow!40}-1.47, 0.81\\
			\cline{3-7}
			&\textbf{C$^{(2)}$} &\cellcolor{green!25}11.49, 13.58 & \cellcolor{green!25}8.18, 10.27 & \cellcolor{green!25}4.20, 6.16 & \cellcolor{green!25}1.53, 3.81 & \cellcolor{yellow!40}-1.48, 0.80\\
			\cline{3-7}
			&\textbf{C$^{(3)}$} &\cellcolor{green!25}19.87, 21.96 & \cellcolor{green!25}16.56, 18.65 & \cellcolor{green!25}12.58, 14.54 & \cellcolor{green!25}9.91, 12.19 & \cellcolor{green!25}6.90, 9.18\\
			\cline{3-7}
			&\textbf{C$^{(4)}$} &\cellcolor{green!25}17.27, 19.36 & \cellcolor{green!25}13.96, 16.05 & \cellcolor{green!25}9.98, 11.94 & \cellcolor{green!25}7.31, 9.59 & \cellcolor{green!25}4.30, 6.58\\
			\cline{3-7}
			&\textbf{C$^{(5)}$} &\cellcolor{green!25}10.64, 12.73 & \cellcolor{green!25}7.33, 9.42 & \cellcolor{green!25}3.35, 5.31 & \cellcolor{green!25}0.68, 2.96 & \cellcolor{red!40}-2.33, -0.05\\
			\cline{3-7}
			&\textbf{D$^{(1)}$} &\cellcolor{green!25}13.52, 15.61 & \cellcolor{green!25}10.21, 12.30 & \cellcolor{green!25}6.23, 8.19 & \cellcolor{green!25}3.56, 5.84 & \cellcolor{green!25}0.55, 2.83\\
			\cline{3-7}
			&\textbf{D$^{(2)}$} &\cellcolor{green!25}15.67, 17.76 & \cellcolor{green!25}12.36, 14.45 & \cellcolor{green!25}8.38, 10.34 & \cellcolor{green!25}5.71, 7.99 & \cellcolor{green!25}2.70, 4.98\\
			\cline{3-7}
			&\textbf{D$^{(3)}$} &\cellcolor{green!25}15.15, 17.24 & \cellcolor{green!25}11.84, 13.93 & \cellcolor{green!25}7.86, 9.82 & \cellcolor{green!25}5.19, 7.47 & \cellcolor{green!25}2.18, 4.46\\
			\cline{3-7}
			&\multicolumn{1}{c}{}&\multicolumn{5}{c}{\cellcolor{gray!25}\textbf{INTERORBITAL $\boldsymbol{k}$-to-$\boldsymbol{k}$}}\\
			\cline{3-7}
			\multicolumn{1}{c}{\multirow{13}{*}{\rotatebox{90}{\textbf{Constellation Shell}}}}
			&\textbf{A$^{(1)}$} & \cellcolor{green!25}6.37, 8.46 & \cellcolor{green!25}3.06, 5.15 & \cellcolor{yellow!40}-0.92, 1.04 & \cellcolor{red!40}-3.59, -1.31 & \cellcolor{red!40}-6.60, -4.32\\
			\cline{3-7}
			&\textbf{A$^{(2)}$} &\cellcolor{green!25}9.87, 11.96 & \cellcolor{green!25}6.56, 8.65 & \cellcolor{green!25}2.58, 4.54 & \cellcolor{yellow!40}-0.09, 2.19 & \cellcolor{red!40}-3.1, -0.82\\
			\cline{3-7}
			&\textbf{B$^{(1)}$} &\cellcolor{green!25}13.04, 15.13 & \cellcolor{green!25}9.73, 11.82 & \cellcolor{green!25}5.75, 7.71 & \cellcolor{green!25}3.08, 5.36 & \cellcolor{green!25}0.07, 2.35\\
			\cline{3-7}
			&\textbf{B$^{(2)}$} &\cellcolor{green!25}11.71, 13.80 & \cellcolor{green!25}8.40, 10.49 & \cellcolor{green!25}4.42, 6.38 & \cellcolor{green!25}1.75, 4.03 & \cellcolor{yellow!40}-1.26, 1.02\\
			\cline{3-7}
			&\textbf{B$^{(3)}$} &\cellcolor{green!25}11.30, 13.39 & \cellcolor{green!25}7.99, 10.08 & \cellcolor{green!25}4.01, 5.97 & \cellcolor{green!25}1.34, 3.62 & \cellcolor{yellow!40}-1.67, 0.61\\
			\cline{3-7}
			&\textbf{C$^{(1)}$} &\cellcolor{green!25}9.97, 12.06 & \cellcolor{green!25}6.66, 8.75 & \cellcolor{green!25}2.68, 4.64 & \cellcolor{green!25}0.01, 2.29 & \cellcolor{red!40}-3.00, -0.72\\
			\cline{3-7}
			&\textbf{C$^{(2)}$} &\cellcolor{green!25}9.95, 12.04 & \cellcolor{green!25}6.64, 8.73 & \cellcolor{green!25}2.66, 4.62 & \cellcolor{yellow!40}-0.01, 2.27 & \cellcolor{red!40}-3.02, -0.74\\
			\cline{3-7}
			&\textbf{C$^{(3)}$} &\cellcolor{green!25}3.28, 5.37 & \cellcolor{yellow!40}-0.03, 2.06 & \cellcolor{red!40}-4.01, -2.05 & \cellcolor{red!40}-6.68, -4.40 & \cellcolor{red!40}-9.69, -7.41\\
			\cline{3-7}
			&\textbf{C$^{(4)}$} &\cellcolor{green!25}3.28, 5.37 & \cellcolor{yellow!40}-0.03, 2.06 & \cellcolor{red!40}-4.01, -2.05 & \cellcolor{red!40}-6.68, -4.40 & \cellcolor{red!40}-9.69, -7.41\\
			\cline{3-7}
			&\textbf{C$^{(5)}$} &\cellcolor{green!25}8.57, 10.66 & \cellcolor{green!25}5.26, 7.35 & \cellcolor{green!25}1.28, 3.24 & \cellcolor{yellow!40}-1.39, 0.89 & \cellcolor{red!40}-4.40, -2.12\\
			\cline{3-7}
			&\textbf{D$^{(1)}$} &\cellcolor{green!25}8.07, 10.16 & \cellcolor{green!25}4.76, 6.85 & \cellcolor{green!25}0.78, 2.74 & \cellcolor{yellow!40}-1.89, 0.39 & \cellcolor{red!40}-4.90, -2.62\\
			\cline{3-7}
			&\textbf{D$^{(2)}$} &\cellcolor{green!25}10.40, 12.49 & \cellcolor{green!25}7.09, 9.18 & \cellcolor{green!25}3.11, 5.07 & \cellcolor{green!25}0.44, 2.72 & \cellcolor{red!40}-2.57, -0.29\\
			\cline{3-7}
			&\textbf{D$^{(3)}$} &\cellcolor{green!25}10.21, 12.30 & \cellcolor{green!25}6.90, 8.99 & \cellcolor{green!25}2.92, 4.88 & \cellcolor{green!25}0.25, 2.53 & \cellcolor{red!40}-2.76, -0.48\\
			\cline{3-7}
			&\multicolumn{1}{c}{}&\multicolumn{5}{c}{\cellcolor{gray!25}\textbf{INTERORBITAL $\boldsymbol{k}$-to-$\boldsymbol{k-1}$}}\\
			\cline{3-7}
			\multicolumn{1}{c}{\multirow{13}{*}{\rotatebox{90}{\textbf{ }}}}
			&\textbf{A$^{(1)}$} & \cellcolor{green!25}5.05, 7.14 & \cellcolor{green!25}1.74, 3.83 & \cellcolor{red!40}-2.24, -0.28 & \cellcolor{red!40}-4.91, -2.63 & \cellcolor{red!40}-7.92, -5.64\\
			\cline{3-7}
			&\textbf{A$^{(2)}$} &\cellcolor{green!25}15.9, 17.99 & \cellcolor{green!25}12.59, 14.68 & \cellcolor{green!25}8.61, 10.57 & \cellcolor{green!25}5.94, 8.22 & \cellcolor{green!25}2.93, 5.21\\
			\cline{3-7}
			&\textbf{B$^{(1)}$} &\cellcolor{green!25}13.26, 15.35 & \cellcolor{green!25}9.95, 12.04 & \cellcolor{green!25}5.97, 7.93 & \cellcolor{green!25}3.3, 5.58 & \cellcolor{green!25}0.29, 2.57\\
			\cline{3-7}
			&\textbf{B$^{(2)}$} &\cellcolor{green!25}14.02, 16.11 & \cellcolor{green!25}10.71, 12.80 & \cellcolor{green!25}6.73, 8.69 & \cellcolor{green!25}4.06, 6.34 & \cellcolor{green!25}1.05, 3.33\\
			\cline{3-7}
			&\textbf{B$^{(3)}$} &\cellcolor{green!25}14.04, 16.13 & \cellcolor{green!25}10.73, 12.82 & \cellcolor{green!25}6.75, 8.71 & \cellcolor{green!25}4.08, 6.36 & \cellcolor{green!25}1.07, 3.35\\
			\cline{3-7}
			&\textbf{C$^{(1)}$} &\cellcolor{green!25}12.88, 14.97 & \cellcolor{green!25}9.57, 11.66 & \cellcolor{green!25}5.59, 7.55 & \cellcolor{green!25}2.92, 5.20 & \cellcolor{yellow!40}-0.09, 2.19\\
			\cline{3-7}
			&\textbf{C$^{(2)}$} &\cellcolor{green!25}12.88, 14.97 & \cellcolor{green!25}9.57, 11.66 & \cellcolor{green!25}5.59, 7.55 & \cellcolor{green!25}2.92, 5.20 & \cellcolor{yellow!40}-0.09, 2.19\\
			\cline{3-7}
			&\textbf{C$^{(3)}$} &\cellcolor{green!25}3.28, 5.37 & \cellcolor{yellow!40}-0.03, 2.06 & \cellcolor{red!40}-4.01, -2.05 & \cellcolor{red!40}-6.68, -4.40 & \cellcolor{red!40}-9.69, -7.41\\
			\cline{3-7}
			&\textbf{C$^{(4)}$} &\cellcolor{green!25}3.28, 5.37 & \cellcolor{yellow!40}-0.03, 2.06 & \cellcolor{red!40}-4.01, -2.05 & \cellcolor{red!40}-6.68, -4.40 & \cellcolor{red!40}-9.69, -7.41\\
			\cline{3-7}
			&\textbf{C$^{(5)}$} &\cellcolor{green!25}10.97, 13.06 & \cellcolor{green!25}7.66, 9.75 & \cellcolor{green!25}3.68, 5.64 & \cellcolor{green!25}1.01, 3.29 & \cellcolor{yellow!40}-2.00, 0.28\\
			\cline{3-7}
			&\textbf{D$^{(1)}$} &\cellcolor{green!25}13.78, 15.87 & \cellcolor{green!25}10.47, 12.56 & \cellcolor{green!25}6.49, 8.45 & \cellcolor{green!25}3.82, 6.10 & \cellcolor{green!25}0.81, 3.09\\
			\cline{3-7}
			&\textbf{D$^{(2)}$} &\cellcolor{green!25}15.85, 17.94 & \cellcolor{green!25}12.54, 14.63 & \cellcolor{green!25}8.56, 10.52 & \cellcolor{green!25}5.89, 8.17 & \cellcolor{green!25}2.88, 5.16\\
			\cline{3-7}
			&\textbf{D$^{(3)}$} &\cellcolor{green!25}15.31, 17.40 & \cellcolor{green!25}12.00, 14.09 & \cellcolor{green!25}8.02, 9.98 & \cellcolor{green!25}5.35, 7.63 & \cellcolor{green!25}2.34, 4.62\\
			\cline{3-7}
		\end{tabular}
	\end{table*}
	
	\begin{table*}[ht!]
		\centering
		\caption{Suitability of different modulation formats/data rates for the various FNCs in the ASE-limited regime. The numbers in sequence within each cell indicate the margin, in dB, for using Staircase and oFEC as correction methods, respectively.}
		\label{tab:tabase}
		\begin{tabular}{>{\centering}m{0.01\textwidth}>{\centering}m{0.055\textwidth}|>{\centering}m{0.14\textwidth}|>{\centering}m{0.14\textwidth}|>{\centering}m{0.145\textwidth}|>{\centering}m{0.157\textwidth}|>{\centering\arraybackslash}m{0.157\textwidth}|}
			&\multicolumn{1}{c}{}&\multicolumn{5}{c}{\textbf{Modulation Format/Data Rate}}\\
			&\multicolumn{1}{c}{}&\multicolumn{1}{c}{\textbf{100G QPSK}}
			&\multicolumn{1}{c}{\textbf{200G QPSK}}
			&\multicolumn{1}{c}{\textbf{300G 8-QAM}} & \multicolumn{1}{c}{\textbf{400G 16-QAM}} & \multicolumn{1}{c}{\textbf{800G 16-QAM}}\\
			&\multicolumn{1}{c}{}&\multicolumn{5}{c}{\cellcolor{gray!25}\textbf{INTRAORBITAL}}\\
			\cline{3-7}
			\multicolumn{1}{c}{\multirow{13}{*}{\rotatebox{90}{\textbf{ }}}}
			&\textbf{A$^{(1)}$} & \cellcolor{green!25}7.14, 9.23 & \cellcolor{green!25}3.83, 5.92 & \cellcolor{yellow!40}-0.15, 1.81 & \cellcolor{red!40}-2.82, -0.54 & \cellcolor{red!40}-5.83, -3.55\\
			\cline{3-7}
			&\textbf{A$^{(2)}$} &\cellcolor{green!25}14.8, 16.89 & \cellcolor{green!25}11.49, 13.58 & \cellcolor{green!25}7.51, 9.47 & \cellcolor{green!25}4.84, 7.12 & \cellcolor{green!25}1.83, 4.11\\
			\cline{3-7}
			&\textbf{B$^{(1)}$} &\cellcolor{green!25}18.37, 20.46 & \cellcolor{green!25}15.06, 17.15 & \cellcolor{green!25}11.08, 13.04 & \cellcolor{green!25}8.41, 10.69 & \cellcolor{green!25}5.4, 7.68\\
			\cline{3-7}
			&\textbf{B$^{(2)}$} &\cellcolor{green!25}21.71, 23.8 & \cellcolor{green!25}18.4, 20.49 & \cellcolor{green!25}14.42, 16.38 & \cellcolor{green!25}11.75, 14.03 & \cellcolor{green!25}8.74, 11.02\\
			\cline{3-7}
			&\textbf{B$^{(3)}$} &\cellcolor{green!25}21.71, 23.8 & \cellcolor{green!25}18.4, 20.49 & \cellcolor{green!25}14.42, 16.38 & \cellcolor{green!25}11.75, 14.03 & \cellcolor{green!25}8.74, 11.02\\
			\cline{3-7}
			&\textbf{C$^{(1)}$} &\cellcolor{green!25}12.23, 14.32 & \cellcolor{green!25}8.92, 11.01 & \cellcolor{green!25}4.94, 6.90 & \cellcolor{green!25}2.27, 4.55 & \cellcolor{yellow!40}-0.74, 1.54\\
			\cline{3-7}
			&\textbf{C$^{(2)}$} &\cellcolor{green!25}12.22, 14.31 & \cellcolor{green!25}8.91, 11.00 & \cellcolor{green!25}4.93, 6.89 & \cellcolor{green!25}2.26, 4.54 & \cellcolor{yellow!40}-0.75, 1.53\\
			\cline{3-7}
			&\textbf{C$^{(3)}$} &\cellcolor{green!25}20.60, 22.69 & \cellcolor{green!25}17.29, 19.38 & \cellcolor{green!25}13.31, 15.27 & \cellcolor{green!25}10.64, 12.92 & \cellcolor{green!25}7.63, 9.91\\
			\cline{3-7}
			&\textbf{C$^{(4)}$} &\cellcolor{green!25}18.00, 20.09 & \cellcolor{green!25}14.69, 16.78 & \cellcolor{green!25}10.71, 12.67 & \cellcolor{green!25}8.04, 10.32 & \cellcolor{green!25}5.03, 7.31\\
			\cline{3-7}
			&\textbf{C$^{(5)}$} &\cellcolor{green!25}11.37, 13.46 & \cellcolor{green!25}8.06, 10.15 & \cellcolor{green!25}4.08, 6.04 & \cellcolor{green!25}1.41, 3.69 & \cellcolor{yellow!40}-1.60, 0.68\\
			\cline{3-7}
			&\textbf{D$^{(1)}$} &\cellcolor{green!25}14.25, 16.34 & \cellcolor{green!25}10.94, 13.03 & \cellcolor{green!25}6.96, 8.92 & \cellcolor{green!25}4.29, 6.57 & \cellcolor{green!25}1.28, 3.56\\
			\cline{3-7}
			&\textbf{D$^{(2)}$} &\cellcolor{green!25}16.40, 18.49 & \cellcolor{green!25}13.09, 15.18 & \cellcolor{green!25}9.11, 11.07 & \cellcolor{green!25}6.44, 8.72 & \cellcolor{green!25}3.43, 5.71\\
			\cline{3-7}
			&\textbf{D$^{(3)}$} &\cellcolor{green!25}15.88, 17.97 & \cellcolor{green!25}12.57, 14.66 & \cellcolor{green!25}8.59, 10.55 & \cellcolor{green!25}5.92, 8.20 & \cellcolor{green!25}2.91, 5.19\\
			\cline{3-7}
			&\multicolumn{1}{c}{}&\multicolumn{5}{c}{\cellcolor{gray!25}\textbf{INTERORBITAL $\boldsymbol{k}$-to-$\boldsymbol{k}$}}\\
			\cline{3-7}
			\multicolumn{1}{c}{\multirow{13}{*}{\rotatebox{90}{\textbf{Constellation Shell}}}}
			&\textbf{A$^{(1)}$} & \cellcolor{green!25}7.09, 9.18 & \cellcolor{green!25}3.78, 5.87 & \cellcolor{yellow!40}-0.20, 1.76 & \cellcolor{red!40}-2.87, -0.59 & \cellcolor{red!40}-5.88, -3.60\\
			\cline{3-7}
			&\textbf{A$^{(2)}$} &\cellcolor{green!25}10.60, 12.69 & \cellcolor{green!25}7.29, 9.38 & \cellcolor{green!25}3.31, 5.27 & \cellcolor{green!25}0.64, 2.92 & \cellcolor{red!40}-2.37, -0.09\\
			\cline{3-7}
			&\textbf{B$^{(1)}$} &\cellcolor{green!25}13.77, 15.86 & \cellcolor{green!25}10.46, 12.55 & \cellcolor{green!25}6.48, 8.44 & \cellcolor{green!25}3.81, 6.09 & \cellcolor{green!25}0.80, 3.08\\
			\cline{3-7}
			&\textbf{B$^{(2)}$} &\cellcolor{green!25}12.44, 14.53 & \cellcolor{green!25}9.13, 11.22 & \cellcolor{green!25}5.15, 7.11 & \cellcolor{green!25}2.48, 4.76 & \cellcolor{yellow!40}-0.53, 1.75\\
			\cline{3-7}
			&\textbf{B$^{(3)}$} &\cellcolor{green!25}12.03, 14.12 & \cellcolor{green!25}8.72, 10.81 & \cellcolor{green!25}4.74, 6.7 & \cellcolor{green!25}2.07, 4.35 & \cellcolor{yellow!40}-0.94, 1.34\\
			\cline{3-7}
			&\textbf{C$^{(1)}$} &\cellcolor{green!25}10.7, 12.79 & \cellcolor{green!25}7.39, 9.48 & \cellcolor{green!25}3.41, 5.37 & \cellcolor{green!25}0.74, 3.02 & \cellcolor{yellow!40}-2.27, 0.01\\
			\cline{3-7}
			&\textbf{C$^{(2)}$} & \cellcolor{green!25}10.68, 12.77 & \cellcolor{green!25}7.37, 9.46 & \cellcolor{green!25}3.39, 5.35 & \cellcolor{green!25}0.72, 3.00 & \cellcolor{red!40}-2.29, -0.01\\
			\cline{3-7}
			&\textbf{C$^{(3)}$} &\cellcolor{green!25}4.01, 6.10 & \cellcolor{green!25}0.70, 2.79 & \cellcolor{red!40}-3.28, -1.32 & \cellcolor{red!40}-5.95, -3.67 & \cellcolor{red!40}-8.96, -6.68\\
			\cline{3-7}
			&\textbf{C$^{(4)}$} &\cellcolor{green!25}4.01, 6.10 & \cellcolor{green!25}0.70, 2.79 & \cellcolor{red!40}-3.28, -1.32 & \cellcolor{red!40}-5.95, -3.67 & \cellcolor{red!40}-8.96, -6.68\\
			\cline{3-7}
			&\textbf{C$^{(5)}$} &\cellcolor{green!25}9.29, 11.38 & \cellcolor{green!25}5.98, 8.07 & \cellcolor{green!25}2.00, 3.96 & \cellcolor{yellow!40}-0.67, 1.61 & \cellcolor{red!40}-3.68, -1.40\\
			\cline{3-7}
			&\textbf{D$^{(1)}$} &\cellcolor{green!25}8.80, 10.89 & \cellcolor{green!25}5.49, 7.58 & \cellcolor{green!25}1.51, 3.47 & \cellcolor{yellow!40}-1.16, 1.12 & \cellcolor{red!40}-4.17, -1.89\\
			\cline{3-7}
			&\textbf{D$^{(2)}$} &\cellcolor{green!25}11.13, 13.22 & \cellcolor{green!25}7.82, 9.91 & \cellcolor{green!25}3.84, 5.80 & \cellcolor{green!25}1.17, 3.45 & \cellcolor{yellow!40}-1.84, 0.44\\
			\cline{3-7}
			&\textbf{D$^{(3)}$} &\cellcolor{green!25}10.94, 13.03 & \cellcolor{green!25}7.63, 9.72 & \cellcolor{green!25}3.65, 5.61 & \cellcolor{green!25}0.98, 3.26 & \cellcolor{yellow!40}-2.03, 0.25\\
			\cline{3-7}
			&\multicolumn{1}{c}{}&\multicolumn{5}{c}{\cellcolor{gray!25}\textbf{INTERORBITAL $\boldsymbol{k}$-to-$\boldsymbol{k-1}$}}\\
			\cline{3-7}
			\multicolumn{1}{c}{\multirow{13}{*}{\rotatebox{90}{\textbf{ }}}}
			&\textbf{A$^{(1)}$} & \cellcolor{green!25}5.78, 7.87 & \cellcolor{green!25}2.47, 4.56 & \cellcolor{yellow!40}-1.51, 0.45 & \cellcolor{red!40}-4.18, -1.90 & \cellcolor{red!40}-7.19, -4.91\\
			\cline{3-7}
			&\textbf{A$^{(2)}$} &\cellcolor{green!25}16.63, 18.72 & \cellcolor{green!25}13.32, 15.41 & \cellcolor{green!25}9.34, 11.3 & \cellcolor{green!25}6.67, 8.95 & \cellcolor{green!25}3.66, 5.94\\
			\cline{3-7}
			&\textbf{B$^{(1)}$} &\cellcolor{green!25}13.99, 16.08 & \cellcolor{green!25}10.68, 12.77 & \cellcolor{green!25}6.70, 8.66 & \cellcolor{green!25}4.03, 6.31 & \cellcolor{green!25}1.02, 3.30\\
			\cline{3-7}
			&\textbf{B$^{(2)}$} &\cellcolor{green!25}14.74, 16.83 & \cellcolor{green!25}11.43, 13.52 & \cellcolor{green!25}7.45, 9.41 & \cellcolor{green!25}4.78, 7.06 & \cellcolor{green!25}1.77, 4.05\\
			\cline{3-7}
			&\textbf{B$^{(3)}$} &\cellcolor{green!25}14.77, 16.86 & \cellcolor{green!25}11.46, 13.55 & \cellcolor{green!25}7.48, 9.44 & \cellcolor{green!25}4.81, 7.09 & \cellcolor{green!25}1.80, 4.08\\
			\cline{3-7}
			&\textbf{C$^{(1)}$} &\cellcolor{green!25}13.61, 15.70 & \cellcolor{green!25}10.30, 12.39 & \cellcolor{green!25}6.32, 8.28 & \cellcolor{green!25}3.65, 5.93 & \cellcolor{green!25}0.64, 2.92\\
			\cline{3-7}
			&\textbf{C$^{(2)}$} &\cellcolor{green!25}13.60, 15.69 & \cellcolor{green!25}10.29, 12.38 & \cellcolor{green!25}6.31, 8.27 & \cellcolor{green!25}3.64, 5.92 & \cellcolor{green!25}0.63, 2.91\\
			\cline{3-7}
			&\textbf{C$^{(3)}$} &\cellcolor{green!25}4.01, 6.10 & \cellcolor{green!25}0.70, 2.79 & \cellcolor{red!40}-3.28, -1.32 & \cellcolor{red!40}-5.95, -3.67 & \cellcolor{red!40}-8.96, -6.68\\
			\cline{3-7}
			&\textbf{C$^{(4)}$} &\cellcolor{green!25}4.01, 6.10 & \cellcolor{green!25}0.70, 2.79 & \cellcolor{red!40}-3.28, -1.32 & \cellcolor{red!40}-5.95, -3.67 & \cellcolor{red!40}-8.96, -6.68\\
			\cline{3-7}
			&\textbf{C$^{(5)}$} &\cellcolor{green!25}11.70, 13.79 & \cellcolor{green!25}8.39, 10.48 & \cellcolor{green!25}4.41, 6.37 & \cellcolor{green!25}1.74, 4.02 & \cellcolor{yellow!40}-1.27, 1.01\\
			\cline{3-7}
			&\textbf{D$^{(1)}$} &\cellcolor{green!25}14.51, 16.60 & \cellcolor{green!25}11.20, 13.29 & \cellcolor{green!25}7.22, 9.18 & \cellcolor{green!25}4.55, 6.83 & \cellcolor{green!25}1.54, 3.82\\
			\cline{3-7}
			&\textbf{D$^{(2)}$} &\cellcolor{green!25}16.58, 18.67 & \cellcolor{green!25}13.27, 15.36 & \cellcolor{green!25}9.29, 11.25 & \cellcolor{green!25}6.62, 8.90 & \cellcolor{green!25}3.61, 5.89\\
			\cline{3-7}
			&\textbf{D$^{(3)}$} &\cellcolor{green!25}16.04, 18.13 & \cellcolor{green!25}12.73, 14.82 & \cellcolor{green!25}8.75, 10.71 & \cellcolor{green!25}6.08, 8.36 & \cellcolor{green!25}3.07, 5.35\\
			\cline{3-7}
		\end{tabular}
	\end{table*}
	
	As expected, the SNR margins obtained given the presence of a pre-amplification scheme, characterized by the ASE-limited condition, are superior to those resulting from the shot-noise regime. In many cases, higher-order modulation formats only become accessible in the ASE regime (e.g. 800G 16-QAM for intraorbital connections in C$^{(5)}$ and $k$-to-$k$ interorbital connections in C$^{(1)}$ shells). Furthermore, although the difference between the threshold values for staircase and oFEC is relatively small, many of the evaluated scenarios require the use of oFEC.
	
	Intraorbital connections are in general the least demanding among the evaluated FNCs. The significative SNR margins are a consequence of the high intraorbital densities. Constellations \textbf{B} and \textbf{D} are able to achieve the highest data rate configuration on this topology in both ASE and shot-noise-limited regimes. A$^{(1)}$ polar shell, in turns, allows, in the best scenario, the implementation of 300G 8-QAM, being the architecture that presents the worst performance for this kind of FNC.
	
	In the case of interorbital connections, our analysis shows that the 800G 16-QAM scheme is difficult to implement for $k$-to-$k$-type connections. In the shot-noise-limited regime, only the shells of \textbf{B} constellation are able to establish it, two of them, only with corrections via oFEC. In this topology, the use of a pre-amplification scheme is mandatory for the feasibility of the 400G 8-QAM configuration for the C$^{(1)}$, D$^{(2)}$, and D$^{(3)}$ shells. It is also possible to note that $k$-to-$k-1$-type connections tend to reduce link lengths in medium-inclination orbits when compared to the $k$-to-$k$ ones. This feature enables the use of 800G 16-QAM in all of the \textbf{D} shells given the increase produced in the D$^{(1)}$ SNR margin .
	
	\section{Doppler Shift Characterization}
	
	In this section, the maximum DS amplitude and time derivative for each of the aforementioned systems is computed. The Doppler frequency shift in the absence of relativistic effects, $\Delta f$, is given by \cite{yang2009doppler}:
	\begin{equation}
		\Delta f(t) = \frac{c}{\lambda_s}\left[\frac{d\lVert r(t)\rVert/dt}{c-\dot{r}^d_{ik}\cdot\left(r(t)/\lVert r(t)\rVert\right)}\right],
	\end{equation}
	The relative position vector from the source satellite to the destination one is given by  $r(t)=r_{ik}^d(t)-r_{jl}^s(t)$, where $r_{jl}^s(t)$ is the position of the $l$-th satellite in $j$-th plane, and $r_{ik}^d(t)$ is the position of the $k$-th satellite in the $i$-th plane.
	
	In what follows, the precise values of DS amplitude and time derivative are computed for all constellation architectures described in Table \ref{table:pachler} for both of interorbital connection\footnote{Intraorbital FNCs have null DS due to the inexistence of eccentricity in the WCM trajectories.} topologies ($k$-to-$k$ and $k$-to-$k-1$). Once again, due to the public unavailability of the phase factor parameter, we are assigned to it the value that leads each of the shells to the maximum carrier frequency deviation scenario, $F_\text{max}$. Later, an upper bound is established regarding DS amplitude and time derivative for AACs.
	
	\subsection{First-Neighbor Connections}
	The maximum values for DS and their respective time derivatives in FNCs are numerically assessed using a linear search algorithm. The results are summarized in Table \ref{DSS}. In general, DS peaks for FNCs\footnote{C$^{(3)}$ and C$^{(4)}$ shells do not allow incessant first-neighbor connections.} are relatively small, as are the maximum values presented by their corresponding time derivatives. The highest achieved values,  $\Delta f_\text{max}\approx 4.6$ GHz and $\Delta f_\text{max}\approx 6.3$ GHz, are reached by the C$^{(3)}$ and C$^{(4)}$ polar shells, respectively, both with $\Delta f'_\text{max}\approx 0.1$ GHz/s. For the same shell, topologies $k$-to-$k$ present DS values greater than $k$-to-$k-1$ only in cases of polar orbits (e.g., A$^{(1)} $, B$^{(1)}$, C$^{(3)}$, and C$^{(4)}$), which can be seen as an important guideline in the design of OISLs. It is worth mentioning that, for practical purposes, FNC is the most likely layout to be implemented, since they result in the least possible FSPL among all connection patterns.
	
	\begin{table*}[ht]
		\centering
		\caption{Doppler Shift peaks per shell and their respective time derivative values for first-neighbor interorbit connections assuming $\lambda_s=1550\textrm{ nm}$.
		}
		\label{DSS}
			\begin{tabular}{>{\centering}m{0.07\textwidth}>{\centering}m{0.07\textwidth}c>{\centering}m{0.15\textwidth}>{\centering}m{0.15\textwidth}c>{\centering}m{0.15\textwidth}>{\centering\arraybackslash}m{0.15\textwidth}}
				& 	 & & \multicolumn{2}{c}{$k$-to-$k$-type connection} & & \multicolumn{2}{c}{$k$-to-$k-1$-type connection} \\\cline{4-5}\cline{7-8}
				System & Shell & & $\Delta f_\text{max}$ [GHz] ($F_\text{max}$)
				& $\Delta f_\text{max}'$ [GHz/s] & & $\Delta f_\text{max}$ [GHz] ($F_\text{max}$) & $\Delta f_\text{max}'$ [GHz/s] \\
				\midrule
				\multirow{2}{*}{\textbf{A}} & \cellcolor{blue!10}A$^{(1)}$ & &\cellcolor{blue!10}1.0837 (2) &\cellcolor{blue!10}0.3655 & &\cellcolor{blue!10}0.8607 (26) &\cellcolor{blue!10}0.0046 \\
				& A$^{(2)}$ & & 0.2689 (0) & 0.0006 & & 0.5619 (19) & 0.0885 \\
				\midrule
				\multirow{3}{*}{\textbf{B}} & \cellcolor{blue!10}B$^{(1)}$ & & \cellcolor{blue!10}0.7861 (0) &\cellcolor{blue!10}0.0213 & &\cellcolor{blue!10}0.8121 (34) &\cellcolor{blue!10}0.1919 \\
				& B$^{(2)}$ & & 0.3913 (0) & 0.0010 & & 0.6417 (0) & 0.0045 \\
				& \cellcolor{blue!10}B$^{(3)}$ & & \cellcolor{blue!10}0.2147 (0) &\cellcolor{blue!10}0.0005 & &\cellcolor{blue!10}0.3640 (0) &\cellcolor{blue!10}0.0011\\
				\midrule
				\multirow{5}{*}{\textbf{C}} & C$^{(1)}$ & & 0.1714 (0) & 0.0005 & & 0.3387 (61) & 0.0360 \\
				& \cellcolor{blue!10}C$^{(2)}$ & & \cellcolor{blue!10}0.1700 (0) & \cellcolor{blue!10}0.0005 & & \cellcolor{blue!10}0.3364 (59) & \cellcolor{blue!10}0.0268\\
				& C$^{(3)}$ & & 4.5927 (5) & 0.0974 & & 4.1797 (5) & 0.0356 \\
				& \cellcolor{blue!10}C$^{(4)}$ & & \cellcolor{blue!10}6.3443 (3) & \cellcolor{blue!10}0.0962 &  & \cellcolor{blue!10}5.8956 (3) & \cellcolor{blue!10}0.0494 \\
				& C$^{(5)}$ & & 0.5607 (0) & 0.0024 & & 0.7947 (29) & 0.1142 \\
				\midrule
				\multirow{3}{*}{\textbf{D}} & \cellcolor{blue!10}D$^{(1)}$ & & \cellcolor{blue!10}0.1764 (0) & \cellcolor{blue!10}0.0004 & & \cellcolor{blue!10}0.5766 (4) & \cellcolor{blue!10}0.0202 \\
				& D$^{(2)}$ & & 0.2183 (0) & 0.0005 & & 0.5635 (9) & 0.0685 \\
				& \cellcolor{blue!10}D$^{(3)}$ & & \cellcolor{blue!10}0.3440 (0) & \cellcolor{blue!10}0.0010  & & \cellcolor{blue!10}0.7065 (13) & \cellcolor{blue!10}1.2562 \\
				\bottomrule
			\end{tabular}
	\end{table*}
	
	\subsection{Bounds on the Doppler Shift}
	
	
	An upper bound for the DS magnitude in AACs can be derived by assuming two satellites performing an one-dimensional uniform rectilinear motion (URM) along a line segment of length $2\pi\left(R_\oplus+H\right)$ and traveling in opposite directions. The velocity developed by each satellite in this URM is taken as the tangential velocity in the WCM circular orbit. So we have
	\begin{equation}
	\dot{r}_{ik}^d(H)=-\dot{r}_{jl}^s(H) = \sqrt{\frac{G M_\oplus}{R_\oplus + H}}\hat{t},
	\nonumber
	\end{equation}
	thus
	\begin{equation}
	\lVert\dot{r}(H)\rVert = 2\sqrt{\frac{G M_\oplus}{R_\oplus + H}},
	\label{vr}
	\end{equation}
	where $\hat{t}$ is the unit tangent vector to the orbital path, $G\approx 6.6743\cdot 10^{-11}\text{ m}^3\text{ kg}^{-1}\text{ s}^{-2}$ is the gravitational constant, and $M_\oplus\approx5,972\cdot 10^{24}$ kg is the Earth mass. In the one-dimensional case, the frequency observed, $f^{URM}_d$, can be approximated as
	\begin{equation}
		f_d^\text{URM} \approx\left(1 \pm \frac{\dot{r}}{c} \right)f_s^\text{URM},
		\label{eq:DopplerApprox}
	\end{equation}
	where $f^{URM}_s$ is the emitted frequency. It follows that the frequency deviation is then given by
	\begin{equation}
		\Delta f = \left(\frac{\dot{r}}{c}\right)f_s^\text{URM} \approx \left(\frac{15343}{3\cdot10^8}\right)193.4\cdot10^{12} \approx 10 \textrm{ GHz},
		\nonumber
	\end{equation}
	assuming an orbit at an altitude of $400$ km \cite{Sylvain2020}, i.e., $140$ km lower than the lowest orbit listed in Table \ref{table:pachler}.
	
	The maximum time derivative values exhibited for these systems, even in the worst-case scenarios, do not represent a practical problem for digital coherent receivers. For a typical semiconductor laser, $\Delta f_\textrm{max}'\approx 70$ THz/s \cite{zhang2019impact}, a value that would only be achieved by two satellites traveling in opposite directions, at velocities of $7,590$ m/s (assuming $H=400$ km), and parallel orbits, if the interorbital distance between them is approximately $2$ m, which is unlikely to happen in real-world applications.
	
	The occurrence of DS prompts two main issues to a coherent communication system, both of which are analogous to a frequency offset between receiver and transmitter laser. The first one is the need for accurate wide-range frequency tracking. The second issue is that larger receiver bandwidths are required if frequency tracking is fully accomplished in the digital environment. In this situation, a highly shifted signal may be pushed out of the receiver band, causing the signal to be improperly filtered. In the next section, we present a possible all-digital domain strategy capable of compensating the characterized DS values found in LEO constellations even in the most aggressive AACs scenario.
	
	\section{Doppler-Shift Compensation}
	

	
	In a first-order model approximation, the DS varies linearly over the compensation window, and the instantaneous frequency offset $\Delta f[n]$ is given by
	
	\begin{equation}
		\Delta f[n] = \Delta f_0 + \frac{d\Delta f}{dt} n T_s,
	\end{equation}
	
	\noindent where $\Delta f_0$ is the initial frequency offset,  $d\Delta f/dt$ is the frequency offset derivative, and $T_s$ is the sampling period. This is a reasonable approximation in OISLs, where the symbol time scales are considerably faster than the relative motion between the satellites.  In this case, a discrete-time received signal $y[n]$ can be expressed as
	\begin{equation}
		y[n] = s[n] e^{2 \pi \Delta f[n] n T_s},
		\label{eq:DopplerShiftModel}
	\end{equation}
	where s[n] is the transmitted signal, 
	
	
	A typical frequency recovery algorithm used in digital coherent optical systems is the $M$th-power algorithm \cite{Leven2007}. It first applies the $M$th-power operator to eliminate symbol information and calculates the fast Fourier transform (FFT) of the resulting signal. Finally, the estimated frequency offset is the value that maximizes the FFT, divided by $M$. Mathematically, the $M$th-power algorithm is expressed as
	\begin{equation}
		\widehat{\Delta f} = \frac{1}{M} \textrm{ARGMAX} \left\{ \textrm{FFT}\left[ y[n]^M \right] \right\},
		\label{eq:Mth-power}
	\end{equation}
	
	
	
	and whose compensation limit is
	\begin{align}
	\Delta f < \frac{R_s}{2M}.
	\label{eq:Limit}
	\end{align}
	As carrier frequency estimation (CFE) methods are usually applied at $M = 4$, compensating a  $10$-GHz frequency offset would require an $80$-GBaud receiver, which is higher than the values expected for satellite communications. In the scenarios investigated in this paper, this condition is only met by the 120-GBaud configurations.
	
	To overcome these limitations, we evaluate an existing two-stage compensation method proposed in \cite{Diniz2011}, combining a symmetry-based algorithm as a coarse stage, followed by the $M$th-power as a fine stage. Although the two-stage method has been thoroughly investigated in \cite{Diniz2011}, the assessed scenarios do not apply to those studied in this paper, as described in Section II. Here we evaluate up to $10$-GHz Doppler shifts with linear variation, Nyquist pulse shapes, and excess bandwidths. We also propose a modification to mitigate the effects of the added noise due to the excess bandwidth.

\begin{figure*}
        \centering
        \includegraphics[width=\textwidth]{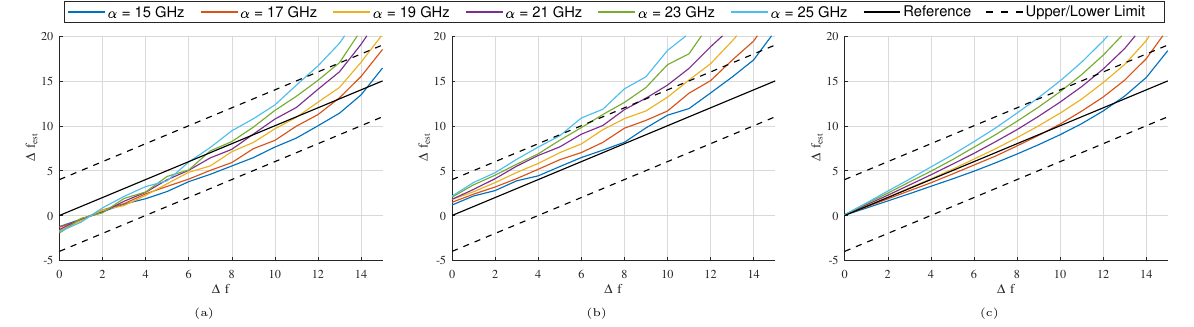}
        \caption{(a) Minimum, (b) maximum, and (c) mean values for different values of the scaling factor $\alpha$. Dashed black lines indicate the upper and lower limits.} \label{fig:scale}
    \end{figure*}

	The coarse CFE algorithm is based on the asymmetry of the received spectrum upon high DSs.
	The DS is estimated as
	\begin{equation}
		f_{est} = \alpha\log\left(\frac{P_+}{P_-} \right),
		\label{eq:Diniz}
	\end{equation}
	where $P_+$ is the power content on positive frequencies and  $P_-$ is the power content on negative frequencies. The ratio between $P_+$ and $P_-$ provides an indication the imprinted frequency shift. The logarithmic operation maps the result to the $[-\infty, \infty]$ range. The scaling factor $\alpha$, which converts the resulting value to frequency, was obtained through a sequential search algorithm. Initially, the value of 21 GHz was selected from \cite{Diniz2011} and settled as a central value. Then, multiple simulations were performed, varying the value of $\alpha$ from 15 GHz to 25 GHz. Minimum, maximum, and mean estimation values were assessed in each of them, as shown in Figs. \ref{fig:scale}a-c, respectively. These results, together with selection criteria from Eq. \ref{eq:Limit}, drive the decision process for $\alpha$. The M-th power algorithm requires the frequency mismatch to be lower than $R_s/(2M)$. Thus, the coarse estimation must be within the range $[\Delta f \pm R_s/(2M)]$ across the entire range of interest for $\Delta f$, $[0, 10]$ GHz. Figs. \ref{fig:scale}a and \ref{fig:scale}b show that a wide range of $\alpha$ satisfy this criterion. As a deciding test, from the reduced set of $\alpha$ values, the one with minimum mean value error at $\Delta f = 10$ GHz was selected, which was 17 GHz, as depicted in Fig. \ref{fig:scale}c.

\begin{figure}[ht]
		\centering
		\subfloat[Evaluated Receiver. \label{2a}]{
			\begin{tikzpicture}[scale = 1.1, node distance=0.8cm, transform shape,rotate = -90]
				\node (block1) [block, rotate = 90] {Optical Front-End};
				\node (block2) [block, rotate = 90, below of=block1] {A/D Converter};
				\node (block3) [block, rotate = 90, below of=block2, yshift=-0.5cm] {Adaptive Equalizer};
				\node (block4) [block, rotate = 90, left of=block3, minimum width = 3cm, xshift=-1.7cm, rotate = -90, text = red, draw = red] {Coarse CFE};
				\node (block5) [block, rotate = 90, below of=block3] {Fine CFE};
				\node (block6) [block, rotate = 90, below of=block5] {Carrier Phase Estimation};
				\node (block7) [block, rotate = 90, below of=block6] {Decision};
				
				\draw [arrow] (-0.6, 0) -- node[anchor=south, xshift = -0.1cm, rotate = 90] {Signal} (block1);
				
				\draw [arrow] (block1.south)++(0, 1.5cm) -- ++(0.2cm, 0);
				\draw [arrow] (block1.south)++(0, -1.5cm) -- ++(0.2cm, 0);
				
				\draw [arrow] (block2.south)++(0, 1.5cm) -- ++(0.7cm, 0);
				\draw [arrow] (block2.south)++(0, -1.5cm) -- ++(0.7cm, 0);
				
				\draw [arrow, draw = red] (block2.south)++(0, 1.5cm) -| ++(0.2cm, -3.7);
				\draw [arrow, draw = red] (block2.south)++(0, -1.5cm) -| ++(0.4cm, -0.7);
				
				\draw [arrow] (block3.south)++(0, 1.5cm) -- ++(0.2cm, 0);
				\draw [arrow] (block3.south)++(0, -1.5cm) -- ++(0.2cm, 0);
				
				\draw [arrow, draw = red] (block4.north) ++(0.8cm, 0) -- (block5);
				
				\draw [arrow] (block5.south)++(0, 1.5cm) -- ++(0.2cm, 0);
				\draw [arrow] (block5.south)++(0, -1.5cm) -- ++(0.2cm, 0);
				
				\draw [arrow] (block6.south)++(0, 1.5cm) -- ++(0.2cm, 0);
				\draw [arrow] (block6.south)++(0, -1.5cm) -- ++(0.2cm, 0);
				
				\draw [arrow] (block7) -- ++(0.6cm, 0);
			\end{tikzpicture}
		}%
		
		\subfloat[Modified Receiver. \label{2b}]{
			\begin{tikzpicture}[scale = 1, node distance=0.8cm, transform shape, rotate = -90]
				\node (block1) [block, rotate = 90] {Optical Front-End};
				\node (block2) [block, rotate = 90, below of=block1] {A/D Converter};
				\node (block3) [block, rotate = 90, below of=block2, draw = red, text = red] {Coarse CFE};
				\node (block4) [block, rotate = 90, below of=block3, draw = red, text = red] {Low-pass Filter};
				\node (block5) [block, rotate = 90, below of=block4] {Adaptive Equalizer};
				\node (block6) [block, rotate = 90, below of=block5] {Fine CFE};
				\node (block7) [block, rotate = 90, below of=block6] {Carrier Phase Estimation};
				\node (block8) [block, rotate = 90, below of=block7] {Decision};
				
				\draw [arrow] (-0.6, 0) -- node[anchor=south, xshift = -0.1cm, rotate = 90] {Signal} (block1);
				
				
				\draw [arrow] (block1.south)++(0, 1.5cm) -- ++(0.2cm, 0);
				\draw [arrow] (block1.south)++(0, -1.5cm) -- ++(0.2cm, 0);
				
				\draw [arrow] (block2.south)++(0, 1.5cm) -- ++(0.2cm, 0);
				\draw [arrow] (block2.south)++(0, -1.5cm) -- ++(0.2cm, 0);
				
				\draw [arrow, draw = red] (block3.south)++(0, 1.5cm) -- ++(0.2cm, 0);
				\draw [arrow, draw = red] (block3.south)++(0, -1.5cm) -- ++(0.2cm, 0);
				
				\draw [arrow] (block4.south)++(0, 1.5cm) -- ++(0.2cm, 0);
				\draw [arrow] (block4.south)++(0, -1.5cm) -- ++(0.2cm, 0);
				
				\draw [arrow] (block5.south)++(0, 1.5cm) -- ++(0.2cm, 0);
				\draw [arrow] (block5.south)++(0, -1.5cm) -- ++(0.2cm, 0);
				
				\draw [arrow] (block6.south)++(0, 1.5cm) -- ++(0.2cm, 0);
				\draw [arrow] (block6.south)++(0, -1.5cm) -- ++(0.2cm, 0);
				
				\draw [arrow] (block7.south)++(0, 1.5cm) -- ++(0.2cm, 0);
				\draw [arrow] (block7.south)++(0, -1.5cm) -- ++(0.2cm, 0);
				
				\draw [arrow] (block8) -- ++(0.6cm, 0);
			\end{tikzpicture}
		}%
		
		\subfloat[Coarse CFE method \label{2c}]{
			\begin{tikzpicture}[scale = 0.775, node distance = 1.4 cm, transform shape]
				\node (Input) [block, minimum width = 1 cm, minimum height = 1 cm] {$ \left | \bullet \right |^2 $};
				\node (FFT) [block, minimum width = 1 cm, minimum height = 1 cm, right of = Input] { FFT };
				\node (P+) [block, minimum width = 1.8 cm, minimum height = 1 cm, right of = FFT, yshift = 0.8 cm, xshift = .5cm] {$ \displaystyle\sum_0^{N/2} X[n] $};
				\node (P-) [block, minimum width = 1 cm, minimum height = 1 cm, right of = FFT, yshift = -0.8 cm, xshift = .5cm] {$ \displaystyle\sum_{-N/2}^0 X[n] $};
				\node (log+) [block, minimum width = 1.6 cm, minimum height = 1.4 cm, right of = P+, xshift = .6cm] {$ \textrm{log}(\bullet) $};
				\node (log-) [block, minimum width = 1.6 cm, minimum height = 1.4 cm, right of = P-, xshift = .6cm] {$ \textrm{log}(\bullet) $};
				\node (Subtr) [operation, right of = FFT, xshift = 4cm] {\huge-};
				\node (multiply) [operation, right of = Subtr, xshift = -0.2cm] {$ \times $};
				
				\draw [arrow] (-.8, 0) -- node[anchor=south, xshift = -0.1cm, rotate = 90] {signal} (Input);
				\draw[arrow] (Input) -- (FFT);
				\draw[arrow] (FFT.east) -| ($(FFT.east) + (0.2, 0)$) |- (P+);
				\draw[arrow] (FFT.east) -| ($(FFT.east) + (0.2, 0)$) |- (P-);
				
				\node (Scaling) [above of = multiply, yshift = -0.2 cm] {$\alpha$};
				\node (Output) [right of = multiply] {$f_{est}$};
				
				\draw[arrow] (P+) -- (log+);
				\draw[arrow] (P-) -- (log-);
				\draw[arrow] (log+) -| (Subtr);
				\draw[arrow] (log-) -| (Subtr);
				\draw[arrow] (Subtr) -- (multiply);
				\draw[arrow] (Scaling) -- (multiply);
				\draw[arrow] (multiply) -- (Output);
			\end{tikzpicture}
		}%
		\caption{The evaluated receiver with coarse CFE stage parallel to the equalization stage is displayed in (a) \cite{Diniz2011}. The modified receiver in (b)  consists of applying the coarse estimation immediately after estimation, filtering the signal to reduce noise via the symmetry based CFE algorithm shown in (c). The red blocks highlight the changes made regarding the original DSP chain.}
		\label{fig:System}
	\end{figure}
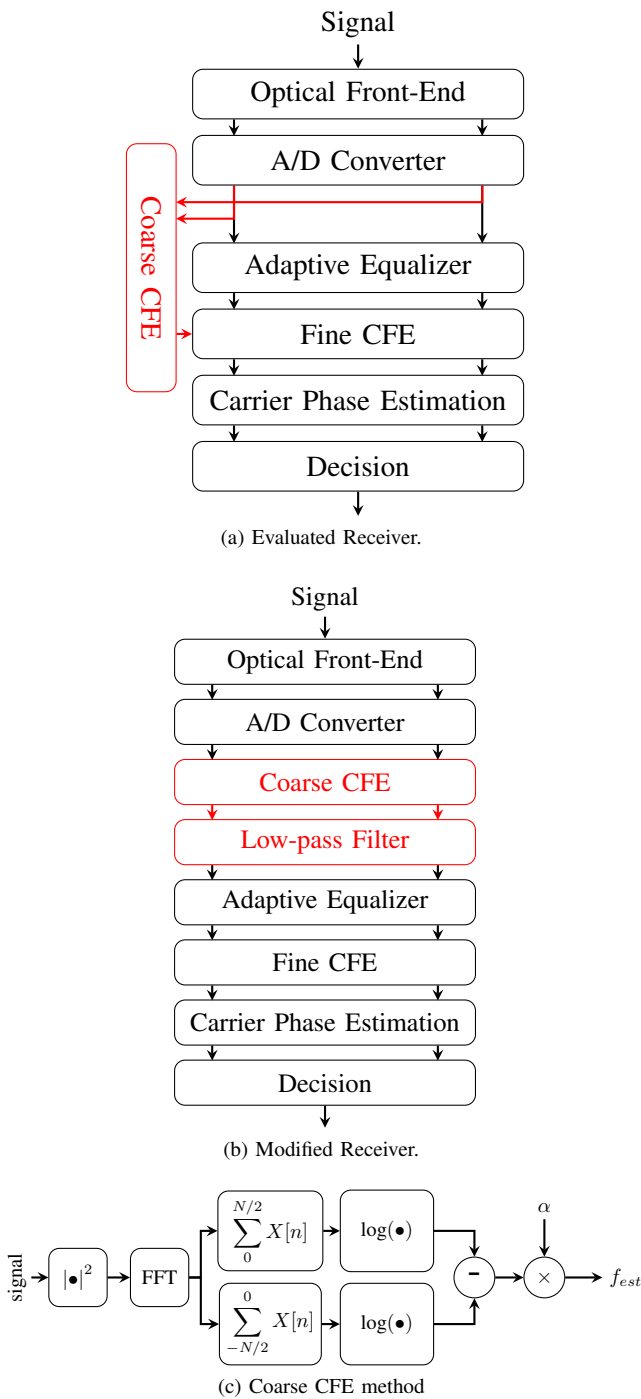

	Fig. \ref{2a} shows the evaluated receiver. An optical front-end model accomplishes the opto-electric conversion process, and an analog-to-digital converter (ADC) model samples the signal at $2$ Sa/Symbol. Coarse frequency estimation and adaptive equalization are carried out in parallel at the sampling rate. Adaptive equalization uses the constant modulus algorithm (CMA) for BPSK and QPSK and the radius directed equalization (RDE) for 16-QAM. After equalization, carrier frequency compensation is performed by the frequency shift estimated through the coarse algorithm. Then, the fine CFE, using the \textit{M}th power algorithm, compensates for the residual frequency offset. Subsequently, carrier phase estimation is attained by the BPS algorithm. Fig. \ref{2b} shows the proposed modified receiver. This system carries the coarse estimation and compensation immediately after sampling. A digital low-pass filter (LPF) filters out excess noise. Then, adaptive equalization, fine CFE, and phase estimation follow similarly to the evaluated receiver. Fig. \ref{2c} details the coarse CFE algorithm, computing the power content on positive and negative frequencies after a FFT.
	
	\section{Simulation Results}
	We assessed excess bandwidth requirements and DS compensation capabilities for the BPSK, QPSK, and 16-QAM modulation formats. Pulse-shaping is carried out by a root-raised-cosine (RRC) filter with a 0.1-roll-off factor. The simulated transmitter and receiver lasers is specified to have a 100-kHz linewidth. The ADC samples the signal at $2$ 
	Sa/Symbol. The coarse CFE uses an 
	$\alpha$ of $17$ GHz and an FFT window of $1024$ samples ($512$ symbols). The CMA (BPSK and QPSK) or RDE (16-QAM) algorithm is implemented with a 21-taps filter. The fine CFE method uses $M = 2$ for BPSK and $M = 4$ for both QPSK and 16-QAM, with an FFT window size of $512$ samples ($512$ symbols). The BPS algorithm uses a 30-symbol window for noise suppression and $40$ test phases. Matching the FFT window symbol size for the coarse and fine CFE algorithms prevents discontinuities in the estimation windows. The simulations assume a 32-GBaud signal with polarization multiplexing and a 1-THz/s frequency time derivative stemming from laser imperfections. A 10th-order super-Gaussian filter generates bandwidth limitations. The modified receiver used a rectangular 40-taps 19.4-GHz bandwidth (roughly $1.1$ times the signal bandwidth) for noise mitigation. The simulations were conducted using $2^{19}$ symbol blocks (per polarization) and differential coding and decoding to address possible cycle slips. The penalties were assessed at a BER of $4\cdot10^{-3}$, slightly lower than the pre-FEC BER threshold for the evaluated staircase FEC of $4.5\cdot10^{-3}$.
	
	
	\begin{figure}
		\centering
		\subfloat[\label{fig:Bandwdith}]{\includegraphics[scale = .6]{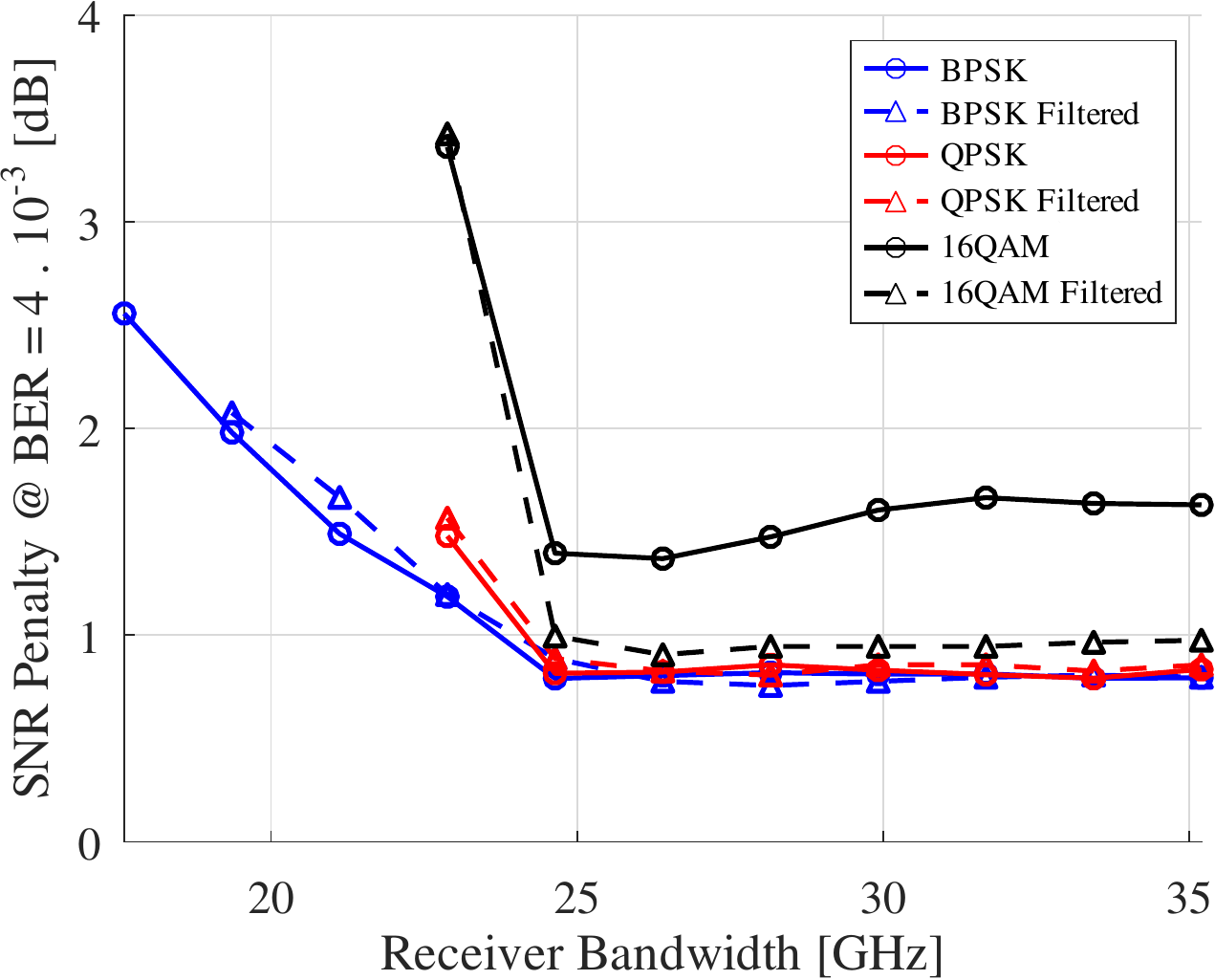}} \\
		\subfloat[\label{fig:Shift}]{\includegraphics[scale = .6]{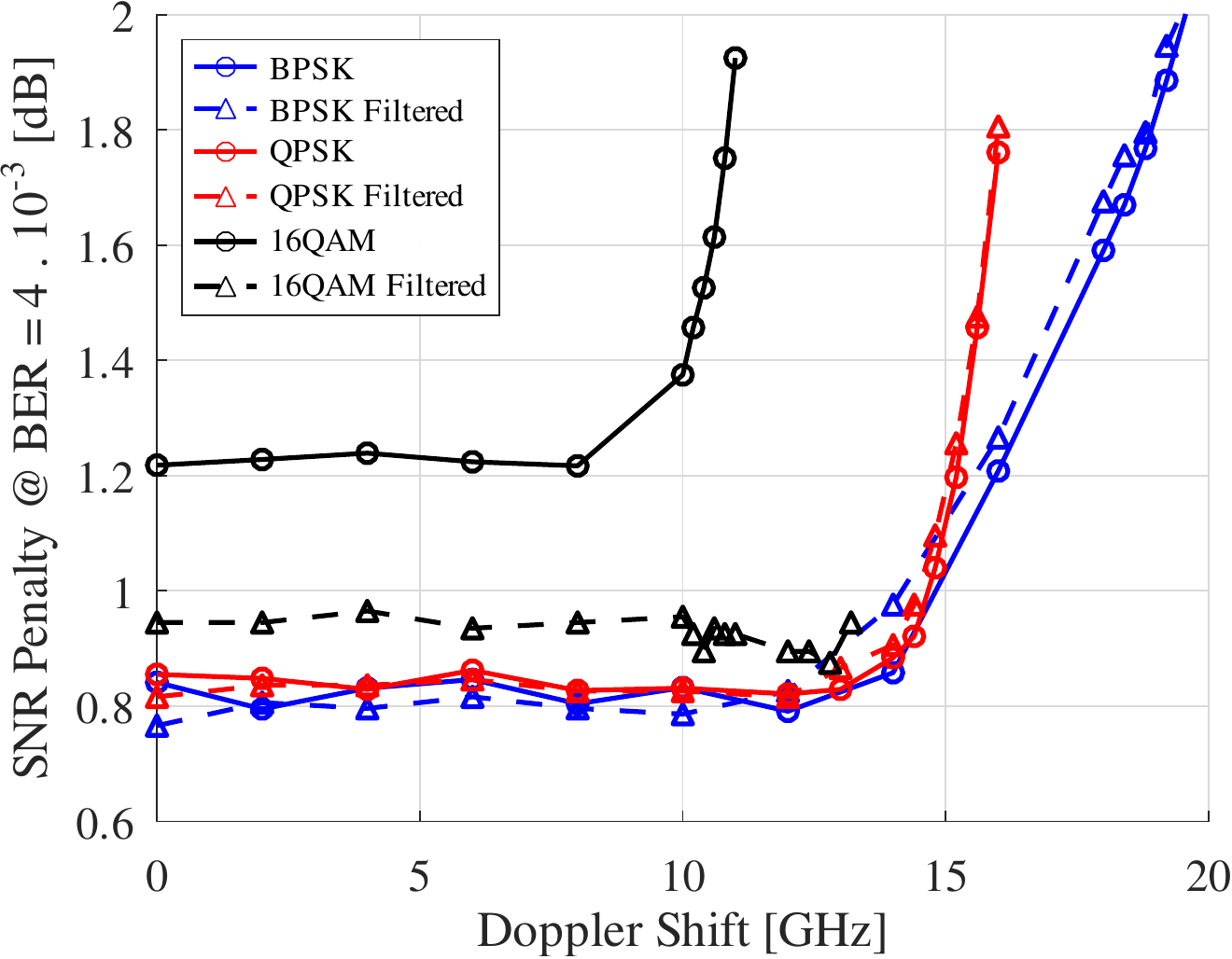}}
		\caption{Comparison between evaluated (in solid lines) and modified receiver (in dashed lines) for each of the modulation formats studied.}
		\label{fig:Simulation}
	\end{figure}
	
	Fig. \ref{fig:Bandwdith} shows the bandwidth required to recover the 17.6-GHz baseband signal added to a 10-GHz Doppler shift. For every scenario, the signal is recovered with a $24$ GHz receiver with a minimum penalty, a slightly smaller value than the $27.6$ GHz expected. This small difference can be explained by the ISI mitigation of the adaptive equalizer. Interestingly, the evaluated receiver with 16-QAM modulation seems more sensitive to the added noise than other scenarios, with penalties increasing with the receiver bandwidth. The modified system reduces the excessive noise mitigating this effect on higher-order modulation formats. For the $10$ GHz shift, the modified receiver reduced the penalties from $1.2$ dB to $0.9$ dB (comparing solid and dashed black lines). Fig. \ref{fig:Simulation}b shows the Doppler compensation limit assuming a $28$-GHz bandwidth for BPSK and QPSK, and $24.5$ GHz for 16-QAM. Although the fine CFE algorithm is limited to $8$ GHz (BPSK), the combination with the coarse CFE stage raises the compensation limit to at least $13$ GHz (16-QAM). The $0.8$ dB penalties come from differential encoding. The 16-QAM penalties include a $0.1$ dB from excessive noise, which was not completely eliminated. 
	
	A comparison between modulation formats shows that higher modulation order causes a more abrupt penalty increase for shifts above $13$ GHz. Both curves (DS and bandwidth) present a similar behavior indicating a relation between available bandwidth and compensation limit. This behavior is also observed in the 16-QAM modified receiver, which has higher bandwidth and larger compensation limit than the 16-QAM original receiver. Both receivers can compensate for shifts of $10$ GHz with controlled penalties meeting the requirements of every constellation.

	\section{Conclusion}
	Modulation formats and DS scenarios for next-generation LEO constellations have been studied assuming different connection topologies. Higher-order modulation formats at high symbol rates can be difficult to implement for interorbital connections in polar shells, given the reduced number of orbital planes. In general, this feature does not hold for intraorbital connections, which, either for polar or medium-inclination shells, typically exhibit ultra-dense orbits. It is also verified that for FNCs both the DS and its time derivative are relatively small and, therefore, amenable to compensation using approaches already available in the literature. However, for the more aggressive scenario of AACs, an alternative methodology for DSC, based on pre-existing techniques, has been proposed, and allows, from a two-stage arrangement, to correct the carrier-frequency deviation in an all-digital domain. BSPK, QPSK, 8-QAM, and 16-QAM modulation formats have been evaluated assuming different symbol rates (28 GBaud, 60 GBaud, and 120 GBaud) and coding schemes (staircase and oFEC). Results regarding bandwidth requirements and compensation capabilities have been presented. Although post-compensation methods avoid transmitter complexity, they have the drawback of requiring extended receiver bandwidths. This should not be an issue in satellite communications, where symbol rates are lower than in typical fiber-optic communication systems.
	
	\bibliography{IEEEabrv,bibfile}
	
\end{document}